\documentclass[12pt,a4paper]{article}
\usepackage{amsmath,amssymb,graphicx,slashed,tabularx}

\allowdisplaybreaks

\newcommand{\TeV}{\,{\rm TeV}}
\newcommand{\GeV}{\,{\rm GeV}}

\newcommand{\invfb}{\,{\rm fb^{-1}}}

\renewcommand{\Re}{\operatorname{Re}}

\makeatletter
\newcommand{\@authornote}[2]{{\def\thefootnote{\fnsymbol{footnote}}\setcounter{footnote}{#1}#2\setcounter{footnote}{0}}}
\newcommand{\authornotemark}[1]{\@authornote#1{\addtocounter{footnote}{-1}\footnotemark}}
\newcommand{\authornotetext}[2]{\@authornote#1{\footnotetext{#2}}}
\makeatother

\begin{document}

\renewcommand{\thefootnote}{\fnsymbol{footnote}} 
\begin{titlepage}

\begin{center}

\hfill UT-15-02\\
\hfill IPMU-15-0018\\
\hfill KEK-TH-1795\\
\hfill February, 2015\\

\vskip .75in

{\Large\bf
Footprints of Supersymmetry on Higgs Decay
}

\vskip .75in

{\large
  Motoi Endo$^{\rm (a,b)}$, Takeo Moroi$^{\rm (a,b)}$, 
  and Mihoko M. Nojiri$^{\rm (b,c,d)}$
}

\vskip 0.25in

$^{\rm (a)}${\em 
Department of Physics, University of Tokyo, Tokyo 113-0033, Japan}

\vskip 0.1in

$^{\rm (b)}${\em 
Kavli IPMU (WPI), University of Tokyo, Kashiwa, Chiba 277-8583, Japan}

\vskip 0.1in

$^{\rm (c)}${\em 
KEK Theory Center, IPNS, KEK, Tsukuba, Ibaraki 305-0801, Japan}

\vskip 0.1in

$^{\rm (d)}${\em 
The Graduate University of Advanced Studies (Sokendai),\\
Tsukuba, Ibaraki 305-0801, Japan}

\end{center}

\vskip .5in

\begin{abstract}

  Motivated by future collider proposals that aim to measure the
  Higgs properties precisely, we study the partial decay
  widths of the lightest Higgs boson in the minimal supersymmetric
  standard model with an emphasis on the parameter region where all
  superparticles and heavy Higgs bosons are not accessible at the LHC.
  Taking account of phenomenological constraints such as the Higgs
  mass, flavor constraints, vacuum stability, and 
  perturbativity of coupling constants up to the grand unification
  scale, we discuss how large the deviations of the partial decay
  widths from the standard model predictions can be. These constraints
  exclude large fraction of the parameter region where the Higgs
  widths show significant deviation from the standard model
  predictions.  Nevertheless, even if superparticles and the heavy
  Higgses are out of the reach of $14\TeV$ LHC, the deviation
  may be large enough to be observed at future $e^+e^-$ collider
  experiments.
  
\end{abstract}
\end{titlepage}

\setcounter{page}{1}
\renewcommand{\thefootnote}{\#\arabic{footnote}}
\setcounter{footnote}{0}

\section{Introduction}
\setcounter{equation}{0}

The discovery of the Higgs boson at ATLAS and CMS experiments
\cite{Aad:2012tfa, Chatrchyan:2012ufa} made a revolutionary impact on
the field of particle physics.  It not only confirmed the so-called
Higgs mechanism for the electroweak symmetry breaking, but also opened
a new possibility to perform a precise test of the standard model (SM)
by studying the properties of the Higgs boson.  In the SM, the coupling
constants of the Higgs boson with other particles are well understood
using the fact that the masses of quarks, leptons, and weak bosons
originate in the vacuum expectation value (VEV) of the Higgs field,
resulting in the prediction of the partial decay widths of the Higgs 
boson into various particles.

In models with physics beyond the SM (BSM), measurements of the Higgs
couplings provide even exciting possibilities.  In large class of BSM
models, there exist new particles at the electroweak to TeV scale,
which affect the properties of the Higgs boson.  Thus, with the
detailed study of the Higgs properties at collider experiments, we
have a chance to observe a signal of BSM physics.  Such a study will
be one of the major subjects in forthcoming collider experiments, i.e., 
the LHC and future $e^+e^-$ colliders like ILC and TLEP \cite{Dawson:2013bba}.

Low energy supersymmetry (SUSY) is a well-motivated candidate of BSM
physics.  Compared to the SM, the particle content is enlarged in SUSY
models.  Even in the minimal setup, i.e., in the minimal SUSY SM
(MSSM), there exist two Higgs doublets, $H_u$ and $H_d$, as well as
superparticles.  The lightest Higgs boson $h$, which plays the role of
the ``Higgs boson'' discovered by ATLAS and CMS, is a linear
combination of the neutral components of $H_u$ and $H_d$, while there
exist other heavier Higgses. In the case where the mass scales of the
heavier Higgses and the superparticles are high enough, the properties
of $h$ are close to those of the SM Higgs boson.  On the contrary, if
the heavier Higgses or superparticles are relatively light, deviations
of the Higgs properties from the SM predictions may be observed by
future collider experiments.  With the precise measurement of the
partial decay widths (or branching ratios) of the Higgs boson,
information about the heavy Higgses and/or superparticles may be
obtained even if those heavy particles can not be directly discovered.

In this paper, we discuss how low the mass scales of the heavier Higgs
bosons and superparticles should be to observe a deviation.  We
evaluate the partial decay widths of the lightest Higgs boson in the
MSSM, taking account of the following phenomenological constraints:
Higgs mass, flavor constraints of the $B$ mesons, stability of the
electroweak (SM-like) vacuum against the transition to charge and
color breaking (CCB) vacua, and perturbativity of coupling constants
up to a high scale.  These constraints exclude large fraction of the
parameter region giving rise to a significant deviation.  Even so, we
will see that the deviations of the partial widths from the SM
predictions can be of $\mathcal{O}(1)\,\%$ for some of the decay modes
in the parameter region allowed by the above-mentioned constraints.
In particular, the deviations may be large enough to be observed by
future future $e^+e^-$ colliders like ILC and TLEP even if
superparticles are so heavy that they would not be observed at the
LHC.

The organization of this paper is as follows.  In Sec.~\ref{sec:MSSM},
we briefly overview the properties of the Higgs bosons in the MSSM.
We also summarize the phenomenological constraints that are taken into
account in our analysis.  Then, in Sec.~\ref{sec:width}, we calculate
the partial decay widths of the lightest Higgs boson in the MSSM and
discuss how large the deviation from the SM prediction can be.
Sec.~\ref{sec:discussion} is devoted for conclusions and discussion.

\section{MSSM: Brief Overview}
\label{sec:MSSM}
\setcounter{equation}{0}

\subsection{Higgs sector of the MSSM}

We review some of the important properties of the Higgs sector in the
MSSM.  There are two Higgs doublets, $H_u$ and $H_d$.  As the neutral
components acquire VEVs, the electroweak symmetry breaking (EWSB)
occurs.  The ratio of the two Higgs VEVs is parameterized by
$\tan\beta\equiv\langle H_u^0\rangle/\langle H_d^0\rangle$.  Assuming
no CP violation in the Higgs potential, the mass eigenstates are
classified as lighter and heavier CP-even Higgs bosons (denoted as $h$
and $H$, respectively), CP-odd (pseudo-scalar) Higgs $A$, and charged
Higgs $H^\pm$.  In the following, we concentrate on the case where the
masses of the heavier Higgses ($H$, $A$, and $H^\pm$) are much larger
than the electroweak scale.  Then, the lightest Higgs boson $h$ should
be identified as the one observed by the LHC.  On the other hand, the
masses of the heavier Higgses are almost degenerate.  We parameterize
the heavier Higgs masses using the pseudo-scalar mass $m_A$.

At the tree level, the lightest Higgs mass is predicted to be smaller
than the $Z$-boson mass, while it is significantly pushed up by
radiative corrections \cite{Okada:1990vk,Okada:1990gg,Ellis:1990nz,Ellis:1991zd,Haber:1990aw}. 
The mass matrix of the neutral CP-even Higgs bosons is denoted as
\begin{align}
\mathcal{M}_h^2 = 
\begin{bmatrix}
  m_Z^2 \cos^2\beta + m_A^2 \sin^2\beta + \delta M^2_{11} &
  -(m_Z^2 + m_A^2) \cos\beta \sin\beta + \delta M^2_{12} \\
  -(m_Z^2 + m_A^2) \cos\beta \sin\beta + \delta M^2_{12} &
  m_Z^2 \sin^2\beta + m_A^2 \cos^2\beta + \delta M^2_{22}
\end{bmatrix},
\label{eq:HiggsMassMatrix}
\end{align}
where $\delta M^2_{ij}$ represents radiative corrections.  

At the one-loop level, the top-stop contribution dominates the
radiative correction to the lightest Higgs mass, and is
approximated as
\begin{align}
\delta m_h^2 \simeq 
\frac{3 m_t^4}{2 \pi^2 v^2}
\left[
\log \frac{m_{\tilde t}^2}{m_t^2} + 
\frac{X_t^2}{m_{\tilde t}^2} \left( 1 - \frac{X_t^2}{12 m_{\tilde t}^2} \right)
\right],
\end{align}
where $v\simeq246\GeV$ is the SM Higgs VEV,
$m_t$ is the top-quark mass,
$m_{\tilde t}^2 \equiv m_{\tilde t_1} m_{\tilde t_2}$ (with
$m_{\tilde t_1}$ and $m_{\tilde t_2}$ being the lighter and heavier
stop masses, respectively), 
and $X_t = A_t - \mu\cot\beta$ (with $A_t$ and $\mu$ being the 
tri-linear scalar couplings for stop and the SUSY
invariant Higgsino mass parameter, respectively).\footnote
{
  In this paper, we adopt the convention of the SLHA format
  \cite{Skands:2003cj}. 
}
The top-stop contribution can significantly enhance the lightest
Higgs mass.  On the other hand, the bottom-sbottom contribution to
the lightest Higgs mass becomes sizable when the bottom Yukawa
coupling is large. It is likely to decrease the lightest Higgs mass.

When stop masses are $\mathcal{O}(1)\TeV$, there are up to four
solutions for $A_t$ to satisfy the observed value of the Higgs mass
$m_h$, for which we use $m_h = 125.7\GeV$ \cite{PDG}.  Let us call these four
solutions as
\begin{itemize}
\item NS: negative $A_t$ with smaller $|A_t|$,
\item NL: negative $A_t$ with larger $|A_t|$,
\item PS: positive $A_t$ with smaller $|A_t|$,
\item PL: positive $A_t$ with larger $|A_t|$.
\end{itemize}
Assuming universal sfermion masses at the SUSY scale, the value of
$|A_t|$ is typically a few times larger than the stop mass for NL and
PL cases.  Such a large value of $|A_t|$ has significant
phenomenological implications, as we will discuss in the next section.

Since the one-loop correction to the Higgs mass is comparable to the
tree-level value, higher order corrections are necessary to obtain
reliable results. In particular, QCD correction, which appears at the
two-loop level, and a large hierarchy between the SUSY scale and the
electroweak scale require the resummation of the leading and sub-leading
logarithms.  We use {\tt FeynHiggs} 2.10.2 \cite{Heinemeyer:1998yj,Heinemeyer:1998np,Degrassi:2002fi,Frank:2006yh,Hahn:2013ria} for the
precise evaluation of the Higgs masses (as well as the mixing
parameters and the partial decay widths of $h$).

At the tree level, $H_u$ ($H_d$) couples only to up-type quarks
(down-type quarks as well as leptons).  However, this is not the case
once radiative corrections due to superparticles are taken into
account.  The Higgs couplings to bottom quark and tau lepton 
can be subject to sizable corrections even when
SUSY breaking scale is very large. Let us parameterize the effective 
$h\bar{b}b$ and $h\bar{t}t$ vertices including radiative corrections 
as \cite{Hall:1993gn,Hempfling:1993kv,Carena:1994bv,Carena:1998gk,Eberl:1999he,Carena:1999py}
\begin{align}
  -\mathcal{L}_{\rm eff} = \, & 
  y_b \, \epsilon_{ij} \bar b_R H_d^i Q_L^j
  + \Delta y_b \, \bar b_R Q_L^k H_u^{k*}
  +
  y_t \, \epsilon_{ij} \bar t_R Q_L^i H_u^j + 
  \Delta y_t \, \bar t_R Q_L^k H_d^{k*}
  + {\rm h.c.},
\end{align}
where $b_R$, $t_R$, and $Q_L$ are right-handed bottom, right-handed top, 
and third-generation quark-doublets, respectively.  In addition, $i$, $j$ 
and $k$ are $SU(2)_L$ indices, while the color indices are omitted for
simplicity. Here, $\Delta y_b$ and $\Delta y_t$ are non-holomorphic radiative
corrections to the Yukawa coupling constants.\footnote
{ For a detailed treatment of the non-holomorphic corrections, 
  see Refs.~\cite{Hofer:2009xb,Noth:2008tw}.
}

After the electroweak symmetry breaking, the Yukawa couplings are
related to the quark masses as\footnote
{
  These relations hold at the SUSY breaking scale. 
  Thus, the quark masses 
  and Yukawa couplings in the formula should be understood as the running 
  parameters at the scale.
}
\begin{align}
  m_b &= \frac{y_b}{\sqrt{2}} v \cos\beta 
  \left( 1 + \frac{\Delta y_b}{y_b} \tan\beta \right)
  \equiv 
  \frac{y_b}{\sqrt{2}}v \cos\beta (1 + \Delta_b), 
  \label{m_b}
  \\
  m_t &= \frac{y_t}{\sqrt{2}} v \sin\beta 
  \left( 1 + \frac{\Delta y_t}{y_t} \cot\beta \right)
  \equiv 
  \frac{y_t}{\sqrt{2}}v \sin\beta (1 + \Delta_t),
\end{align}
where, at the leading order in the mass-insertion approximation,
$\Delta_f$ is given by
\begin{align}
  \Delta_b &\simeq
  \left[
    \frac{2\alpha_s}{3\pi} M_3 \mu\, I(m_{\tilde b_1}^2,m_{\tilde b_2}^2,M_3^2)
    + \frac{y_t^2}{16\pi^2} \mu A_t\, I(m_{\tilde t_1}^2,m_{\tilde t_2}^2,\mu^2)
  \right] \tan\beta
  , \\
  \Delta_t &\simeq
  \left[
    \frac{2\alpha_s}{3\pi} M_3 \mu\, I(m_{\tilde t_1}^2,m_{\tilde t_2}^2,M_3^2)
    + \frac{y_b^2}{16\pi^2} \mu A_b\, I(m_{\tilde b_1}^2,m_{\tilde b_2}^2,\mu^2)
  \right] \cot\beta,
\end{align}
with $M_3$ being the gluino mass.  The loop integral is defined as
\begin{align}
I(a,b,c) = \frac{ab\ln a/b+bc\ln b/c+ca\ln c/a}{(a-b)(b-c)(a-c)}.
\end{align}
Notice that $\Delta_b$ is enhanced when $\tan\beta$ is large, while
$\Delta_t$ is suppressed by $\cot\beta$.

\begin{figure}[t]
\begin{center}
\includegraphics[scale=1]{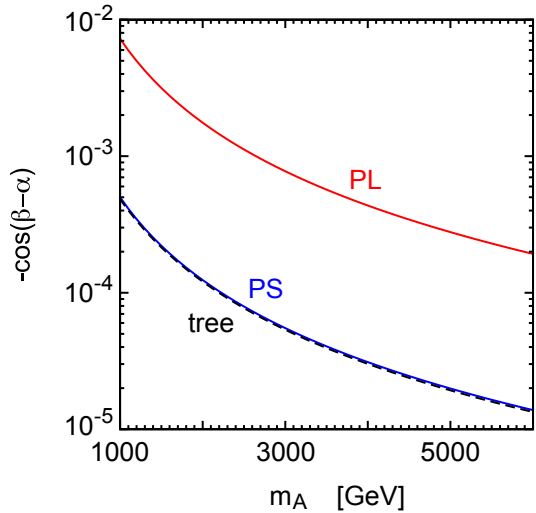}
\end{center}
\caption{$\cos(\beta-\alpha)$ is plotted as a function of $m_A$ with
  $\tan\beta=35$, $M_3=-\mu=5\TeV$, and the approximate GUT relation
  among gaugino masses (see Eq.~\eqref{GUTrelation}).  All the
  sfermion masses are assumed to be universal at $M_{\rm SUSY}$, and
  are also taken to be $5\TeV$. The red and blue lines correspond to
  the cases of the PL and PS solutions of $A_t$, respectively.  The
  black dashed line is the tree-level value.}
\label{fig:cosba}
\end{figure}

The mass eigenstates of the Higgs bosons are given by linear combinations 
of $H_u$ and $H_d$. The CP-even parts of their neutral components are 
related to the mass eigenstates as
\begin{align}
  \Re(H_u^0) & = 
  \frac{1}{\sqrt{2}} (v \sin\beta + h \cos\alpha + H \sin\alpha),
  \\
  \Re(H_d^0) & = 
  \frac{1}{\sqrt{2}} (v \cos\beta - h \sin\alpha + H \cos\alpha).
\end{align}
The mixing angle $\alpha$ depends on the pseudo-scalar mass $m_A$, and
shows a decoupling behaviour, i.e., $\cos(\beta-\alpha)\rightarrow
0$ as $m_A\rightarrow\infty$.  In this limit, $h$ behaves as the SM
Higgs boson.  Using Eq.~\eqref{eq:HiggsMassMatrix}, we obtain
\cite{Carena:2001bg}
\begin{align}
  \cos (\beta-\alpha) 
  = 
  \frac{m_Z^2\sin4\beta}{2m_A^2} 
  \left( 1 + \frac{\delta M^2_{11} - \delta M^2_{22}}{2m_Z^2\cos2\beta} - 
    \frac{\delta M^2_{12}}{m_Z^2\sin2\beta} \right) 
  + \mathcal{O}\left(\frac{m_Z^4}{m_A^4}\right).
\label{eq:cosba}
\end{align}
In Fig.~\ref{fig:cosba}, we show the behavior of $\cos(\beta-\alpha)$
as a function of $m_A$ with the masses of superparticles being fixed.
Here, all the sfermion mass parameters, $m_{\tilde{Q}}$,
$m_{\tilde{U}}$, $m_{\tilde{D}}$, $m_{\tilde{L}}$, and $m_{\tilde{E}}$, 
are taken to be universal at the SUSY scale $M_{\rm SUSY}$, where these 
mass parameters are soft SUSY breaking masses of sfermions with gauge 
quantum numbers of $({\bf 3},{\bf 2},\frac{1}{6})$, $({\bf \bar{3}}, 
{\bf 1}, -\frac{2}{3})$, $({\bf \bar{3}},{\bf 1},\frac{1}{3})$, 
$({\bf 1},{\bf2}, -\frac{1}{2})$, and $({\bf 1},{\bf 1},1)$, respectively, 
with the numbers in the parenthesis being quantum numbers for $SU(3)_C$, 
$SU(2)_L$, and $U(1)_Y$.  Throughout our study, we take $M_{\rm SUSY} = 
(m_{\tilde Q}m_{\tilde U})^{1/2}$.  In addition, the sfermion masses are
assumed to be universal in generation indices.  We can see that radiative 
corrections can enhance $\cos(\beta-\alpha)$ by an order of magnitude when 
$A_t$ is large, while it is comparable to the tree-level value with the 
smaller $|A_t|$ solutions. 

Denoting $h\bar{f}f$ coupling (with $f$ being the SM fermions) as
\begin{align}
  -\mathcal{L}_{h\bar{f}f} \equiv g_{h\bar{f}f} h\bar{f}f,
\end{align}
we obtain the $h\bar{b}b$ coupling constant as
\begin{align}
  g_{h\bar{b}b} & = 
  - \left( \frac{\sin\alpha}{\cos\beta} \right)
  \frac{1 - \Delta_b \cot\alpha \cot\beta}{1 + \Delta_b} 
  g_{h\bar{b}b}^{\rm (SM)}
  \nonumber \\ & = 
  \left[ 
    \sin (\beta-\alpha)
    - \frac{\tan\beta - \Delta_b \cot\beta}{1+\Delta_b} \cos (\beta-\alpha)
  \right]
  g_{h\bar{b}b}^{\rm (SM)},
  \label{g_hbb}
\end{align}
where the superscript ``(SM)'' is used for the SM prediction. When
$m_A$ is relatively large, $\sin(\beta-\alpha)$ is almost unity, while
the second term proportional to $\cos(\beta-\alpha)$ induces a
sizable deviation from the SM value.  Similar relation holds for
$h\bar{\tau}\tau$ vertex, and $\Delta_\tau$ is approximately given by
\begin{align}
  \Delta_\tau &\simeq
  -\frac{3\alpha_2}{8\pi} M_2 \mu \tan\beta\, 
  I(m_{\tilde \tau_L}^2,M_2^2,\mu^2),
  \label{eq:deltatau}
\end{align}
with $M_2$ being the Wino mass. Quantitatively, $\Delta_\tau$ is
smaller than $\Delta_b$ in the parameter space of our study.  As we
will see later, $\Gamma(h\rightarrow\bar{b}b)$ and
$\Gamma(h\rightarrow\bar{\tau}\tau)$ may show sizable deviations from
the SM predictions even if $m_A$ is above TeV.

The $h\bar{t}t$ coupling constant is obtained as
\begin{align}
  g_{h\bar{t}t} & = 
  \left(\frac{\cos\alpha}{\sin\beta}\right)
  \frac{1 - \Delta_t \tan\alpha \tan\beta}{1 + \Delta_t}
  g_{h\bar{t}t}^{\rm (SM)}
  \nonumber \\ & = 
  \left[ 
    \sin (\beta-\alpha)
    + \frac{\cot\beta - \Delta_t \tan\beta}{1+\Delta_t} \cos (\beta-\alpha)
  \right]
  g_{h\bar{t}t}^{\rm (SM)}.
  \label{g_htt}
\end{align}
The deviation from the SM value mainly comes from the second term in
the bracket and is not enhanced by $\tan\beta$, since $\Delta_t$ is 
proportional to $\cot\beta$.

The gauge-boson final states are also important.  In order to
calculate the partial decay widths of the Higgs to gauge bosons, we have
modified {\tt FeynHiggs} 2.10.2 package to properly take account of
the effect of non-holomorphic correction to the Higgs
interaction.\footnote
{ We use Eq.~\eqref{g_hbb} for the $h\bar{b}b$ coupling to calculate the partial 
  decay widths of $h\rightarrow \gamma\gamma$, $\gamma Z$ and $gg$.
  We found that the factor of $(1 - \Delta_b \cot\alpha \cot\beta)$  
  was missing in {\tt FeynHiggs} 2.10.2 package (see Eq.~\eqref{g_hbb}).}
We have also modified the package to take the $\alpha_{\rm eff}$
approximation~\cite{Heinemeyer:2000fa} for calculating the partial widths, 
in which the renormalization scale of the Higgs wave functions is set to 
be $p^2=0$, and the effects
of radiative corrections are included in the mixing between the light
and heavy Higgs bosons.  Then, as we will see below, the partial decay 
widths show proper decoupling behavior in the large $m_A$ limit.\footnote
{ In particular, the partial width $\Gamma(h\rightarrow gg)$ converges to 
  the SM value in the limit of large $m_A$ and squark masses.  This behavior
  looks inconsistent with the result shown in Ref.~\cite{Cahill-Rowley:2014wba}.}

The processes induced by triangle loops, $h\rightarrow gg$ and
$\gamma\gamma$, have been important for the Higgs discovery at the LHC. 
They are also important in studying new particles that couple to the Higgs
boson.  In SUSY models, the stop and sbottom loops contribute to the
$hgg$ coupling.  It is expressed by an approximate formula 
(cf.~Refs.~\cite{Arvanitaki:2011ck,Reece:2012gi,Bagnaschi:2014zla})
\begin{equation}
\frac{g_{hgg}}{g^{\rm (SM)}_{hgg}} \simeq 
\frac{g_{h\bar{t}t}}{g_{h\bar{t}t}^{\rm (SM)}} + 
\sum_{f=t,b}
\frac{m_f^2}{4(1+\Delta_f)^2}
\left(\frac{1}{m^2_{\tilde{f}_1}} + 
\frac{1}{m^2_{\tilde{f}_2}} - 
\frac{X^2_f}{m^2_{\tilde{f}_1}m^2_{\tilde{f}_2}}
\right),
\end{equation}
up to $D$-term and bottom-loop contributions. Here, $X_b=A_b-\mu\tan\beta$.
In the right-hand side, the first term comes from the top loop, while the
second term is given by the stop and sbottom loops. The correction is 
positive in the non-mixing limit (i.e., $X_f\rightarrow 0$), whereas it 
becomes negative when the mixing terms are sizable.  We also
note here that $hVV$ couplings (with $VV=W^+W^-$ and $ZZ$) are
approximately given by $g_{hVV}/g_{hVV}^{\rm (SM)} = \sin
(\beta-\alpha)$. This ratio is very close to unity, and hence the
deviations in these modes are very small.

\subsection{Constraints} \label{sec:constraint}

Before discussing the possibility of observing a deviation of the
Higgs partial widths from the SM prediction at future colliders, we
summarize phenomenological constraints on the MSSM parameter space
which are adopted in our analysis.

\subsubsection{$B_s \to \mu^+\mu^-$}

In the SM, the flavor-changing decay, $B_s \to \mu^+\mu^-$, proceeds
by virtual exchanges of the $Z$ and $W$ bosons.  They are suppressed
by the final-state helicity.  In contrast, SUSY contributions can be
enhanced considerably by large $\tan\beta$, when virtual exchanges of
the heavy Higgs boson contribute to the decay
\cite{Babu:1999hn,Choudhury:1998ze}.  The branching ratio is expressed
as \cite{Buchalla:1993bv,Misiak:1999yg,Bobeth:2001sq,Bobeth:2001jm}
\begin{align}
{\rm Br}(B_s \to \mu^+\mu^-) =  &
\frac{G_F^2 \alpha^2}{64\pi^3} 
f_{B_s}^2 m_{B_s}^3 \tau_{B_s} \left| V_{tb} V_{ts}^* \right|^2 
\sqrt{1-\frac{4m_\mu^2}{m_{B_s}^2}} \nonumber \\
& \times
\left[
\left(1-\frac{4m_\mu^2}{m_{B_s}^2}\right) \left|C_{Q1}-C_{Q1}'\right|^2 + 
\left|\left(C_{Q2}-C_{Q2}'\right) + \frac{2m_\mu}{m_{B_s}} \left(C_{10}-C_{10}'\right) 
\right|^2
\right],
\end{align}
where $G_F$ is the Fermi constant, $m_{B_s}$ is the $B_s$-meson mass,
$f_{B_s}$ is the decay constant of $B_s$, $m_\mu$ is the muon mass,
$\tau_{B_s}$ is the lifetime of $B_s$, and $V_{ij}$ is
Cabbino-Kobayashi-Maskawa matrix element.  In the above expression,
$C_{10}$, $C_{Q1}$ and $C_{Q2}$ are the Wilson coefficients of
the effective operators, $\mathcal{O}_i\propto(\bar s\gamma_\mu P_L
b)(\bar\ell\gamma^\mu\gamma_5\ell)$, $(\bar sP_Rb)(\bar\ell\ell)$, and
$(\bar sP_Rb)(\bar\ell\gamma_5\ell)$, respectively, while $C_i'$ are
obtained by flipping chiralities, $R \leftrightarrow L$.  Among them,
the SM contribution appears only in $C_{10}$.  Including higher order
contributions and taking account of effects of the $B_s\text{-}\bar
B_s$ oscillation, the SM prediction becomes \cite{Bobeth:2013uxa}
\begin{align}
{\rm Br}(B_s \to \mu^+\mu^-)_{\rm SM} = (3.65 \pm 0.23) \times 10^{-9}.
\end{align}
This can be compared with the LHCb measurements \cite{CMS:2014xfa}
\begin{align}
{\rm Br}(B_s \to \mu^+\mu^-)_{\rm LHCb} = (2.8^{+0.7}_{-0.6}) \times 10^{-9}.
\end{align}
Defining $\Delta {\rm Br}(B_s \to \mu^+\mu^-) \equiv 
{\rm Br}(B_s \to \mu^+\mu^-) - {\rm Br}(B_s \to \mu^+\mu^-)_{\rm SM}$,
where the first term in the right-hand side includes both the SUSY and SM contributions,
the 95\% C.L.~bound is estimated as
\begin{align}
-2.3 \times 10^{-9} < \Delta {\rm Br}(B_s \to \mu^+\mu^-) < 0.6 \times 10^{-9}.
\label{br_b2ss}
\end{align}
We will adopt this constraint in our numerical analysis.

Even when superparticles are heavy, they affect the branching ratio
through non-holomorphic contributions to the heavy Higgs couplings.
Including radiative corrections and diagonalizing the quark mass
matrices, effective couplings of the heavy Higgs bosons to the
down-type fermions become \cite{Babu:1999hn,Choudhury:1998ze}
\begin{align}
\mathcal{L}_{\rm eff} \simeq &
\frac{gm_b}{\sqrt{2}m_W\cos\beta} 
\frac{\Delta_{\rm FC}}{(1+\Delta_b)(1+\Delta_0)}
V_{tb}V_{ts}^*\,
(\bar s_L b_R) \left(H + i A\right) \notag\\
& + 
\frac{gm_\ell}{\sqrt{2}m_W\cos\beta} 
\frac{1}{1+\Delta_\ell}
(\bar \ell_L \ell_R) \left(H + i A\right)
+ {\rm h.c.},
\end{align}
where $\Delta_0 = \Delta_b-\Delta_{\rm FC}$. The flavor-changing
coupling is induced by $\Delta_{\rm FC}$ as
\begin{align}
\Delta_{\rm FC} = 
\frac{y_t^2}{16\pi^2} \mu A_t \tan\beta\,
I(m_{\tilde t_1}^2,m_{\tilde t_2}^2,\mu^2).
\end{align}
Here, soft scalar masses are assumed to be universal in generation,
and $\Delta_\ell$ is obtained by substituting $\tilde\tau \rightarrow
\tilde\ell$ in Eq.~\eqref{eq:deltatau}.  Then, the Wilson coefficients
receive Higgs-mediated contributions,
\begin{align}
C_{Q1} \simeq -C_{Q2} \simeq - 
\frac{m_t^2 m_b m_\mu}{4\sin^2\theta_W m_W^2m_A^2} 
\frac{\tan^3\beta}{(1+\Delta_b)^2}
\frac{\mu A_t}{m_{\tilde t}^2}
x_{\tilde t\mu} I (x_{\tilde t\mu},x_{\tilde t\mu},1),
\end{align}
where $m_b$ is the bottom-quark mass, $\theta_W$ is the Weinberg angle,
and $x_{\tilde t\mu}\equiv m_{\tilde t}^2/\mu^2$.
The non-holomorphic correction $\Delta_{\rm FC}$ as well as $\Delta_b$
does not decouple even for very heavy superparticles.  We will see
later that the corrections to ${\rm Br}(B_s \to \mu^+\mu^-)$ can be sizable 
and that the constraint excludes some part of the parameter space of our interest.

The branching ratio of the inclusive decay of $b \to s\gamma$ may also
be sensitive to the non-decoupling contributions to the (charged) Higgs 
boson.  In the numerical analysis, defining $\Delta {\rm Br}(b \to s\gamma) 
\equiv {\rm Br}(b \to s\gamma) - {\rm Br}(b \to s\gamma)_{\rm SM}$, we 
adopt the 95\% C.L.~bound,
\begin{align}
-3.6 \times 10^{-5} < \Delta{\rm Br}(b \to s\gamma) < 9.2 \times 10^{-5}.
\label{eq;bsg}
\end{align}
where the experimental value ${\rm Br}(b \to s\gamma)_{\rm exp}=(3.43 \pm
0.21 \pm 0.07) \times 10^{-4}$ \cite{Amhis:2014hma} and the SM prediction 
${\rm Br}(b \to s\gamma)_{\rm SM}=(3.15 \pm 0.23) \times 10^{-4}$ 
\cite{Misiak:2006zs} are combined. At the current accuracies, $B_s \to 
\mu^+\mu^-$ imposes more stringent bound on the parameter space than 
$b \to s\gamma$ except when superparticles are light.

In the numerical analysis, {\tt SuperIso} 3.4 \cite{Mahmoudi:2008tp} is
used for evaluating the SUSY contributions to the branching ratios as
well as the SM predictions.  In our analysis, we assume that the squark
masses are universal in generations.  If the squark masses are
non-universal, there are extra contributions to $\Delta_{\rm FC}$, and 
the flavor constraints are affected (see e.g., Ref.~\cite{Altmannshofer:2012ks}).
Such a non-universality is expected even in the model with the
universal scalar masses at the GUT scale.  Thus, it should be noted
that the flavor constraints that we will show below are just for a
particular choice of the squark-mass parameters and may change if the
squark mass matrices have non-universal structures.

\subsubsection{Vacuum stability}

With sufficiently large $|A_t|$, CCB vacua arise, and the minimum of
the scalar potential with the correct EWSB becomes a false vacuum
\cite{Frere:1983ag,Gunion:1987qv,Casas:1995pd,Kusenko:1996jn}.  When
$|\mu| \ll |A_t|$, stop and the up-type Higgs fields acquire large
VEVs at the CCB vacua, while the VEVs of other fields are relatively
small.  Recently, the decay rate of the SM-like vacuum has been
studied in detail for such a case
\cite{Camargo-Molina:2013sta,Chowdhury:2013dka,Blinov:2013fta,Camargo-Molina:2014pwa}.
On the other hand, if $\mu$ is as large as the stop masses, the
down-type Higgs boson also has a large VEV at the CCB vacua due to the
tri-linear scalar coupling among stops and the down-type Higgs, which
is proportional to $y_t \mu$. The vacuum stability condition is
important in such a case because significant deviations of the Higgs
partial widths from the SM prediction may occur. In order to study the
SM-like vacuum stability, we consider the tree-level scalar potential
in the field space involving $\tilde{t}_L$, $\tilde{t}_R$, $h_u$ and
$h_d$ (which are canonically normalized scalar fields embedded in the
left-handed stop, right-handed stop, the up-type Higgs and the
down-type Higgs, respectively).

The relevant part of the potential is given by
\begin{align}
V =& \frac{1}{2} m_{11}^2 \, h_d^2 + \frac{1}{2} m_{22}^2 \, h_u^2 - m_{12}^2 \, h_d h_u
+ \frac{1}{2} m_{\tilde Q}^2 \,\tilde t_L^2 + \frac{1}{2} m_{\tilde U}^2 \,\tilde t_R^2 
\notag \\
& + \frac{1}{\sqrt{2}} y_t (A_t h_u - \mu h_d)\tilde t_L \tilde t_R 
+ \frac{1}{4} y_t^2 (\tilde t_L^2 \tilde t_R^2 + \tilde t_L^2 h_u^2 + \tilde t_R^2 h_u^2) 
\notag \\
& + \frac{1}{24} g_3^2 (\tilde t_L^2 - \tilde t_R^2)^2
+ \frac{1}{32} g_2^2 (h_u^2 - h_d^2 - \tilde t_L^2)^2
+ \frac{1}{32} g_Y^2 
\left(h_u^2 - h_d^2 + \frac{1}{3}\tilde t_L^2 - \frac{4}{3}\tilde t_R^2 \right)^2,
\label{eq:stophiggs}
\end{align}
where
\begin{align}
m_{11}^2 &= m_A^2 \sin^2\beta - \frac{1}{2} m_Z^2 \cos 2\beta, \\
m_{22}^2 &= m_A^2 \cos^2\beta + \frac{1}{2} m_Z^2 \cos 2\beta, \\
m_{12}^2 &= \frac{1}{2} m_A^2 \sin 2\beta.
\end{align}

When the CCB vacua become deeper than the SM-like vacuum, the latter
is not stable.  The vacuum decay rate per unit volume is calculated in
the semi-classical approximation, and is expressed as 
\begin{align}
\Gamma/V = C \exp (-S_E),
\end{align}
where $S_E$ is the Euclidean action of the so-called bounce solution
\cite{Coleman:1977py, Callan:1977pt}.  For the decay of the SM-like
vacuum, $C$ is estimated to be $\sim (100\GeV)^4$.  In order for the
lifetime of the SM-like vacuum to be longer than the present age of 
the universe, the Euclidean action is required to satisfy
\begin{align}
  S_E \gtrsim 400.
  \label{S_E>400}
\end{align}
In our numerical analysis, {\tt CosmoTransition} 2.0a1
\cite{Wainwright:2011kj} is used to find the bounce solution in the
four-dimensional field space parameterized by $h_d$, $h_u$, $\tilde
t_L$ and $\tilde t_R$, and to calculate the Euclidean bounce action
$S_E$.  The calculation is done at zero temperature, and therefore
thermal effects are not taken into account.  The model parameters
such as the tri-linear and top Yukawa couplings are evaluated at the
SUSY scale $M_{\rm SUSY}$.

\begin{figure}
\begin{center}
\includegraphics[scale=1.]{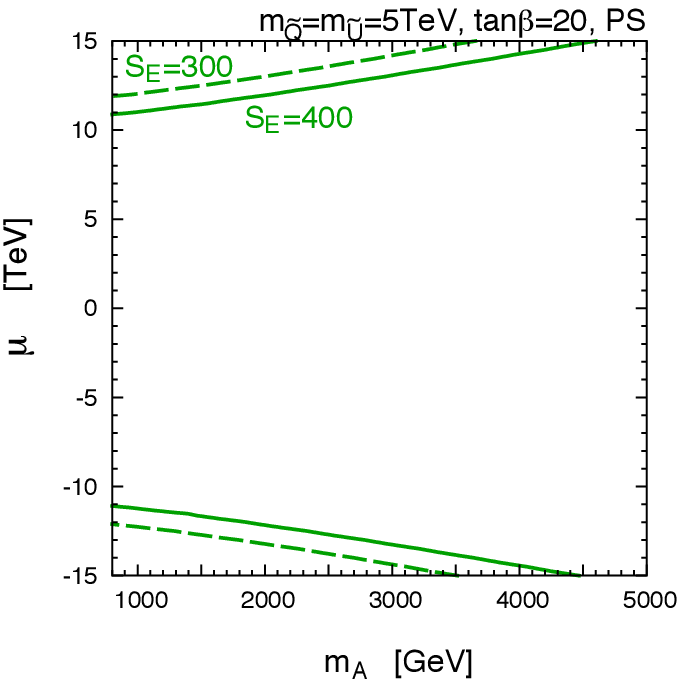} \hspace{5mm}
\includegraphics[scale=1.]{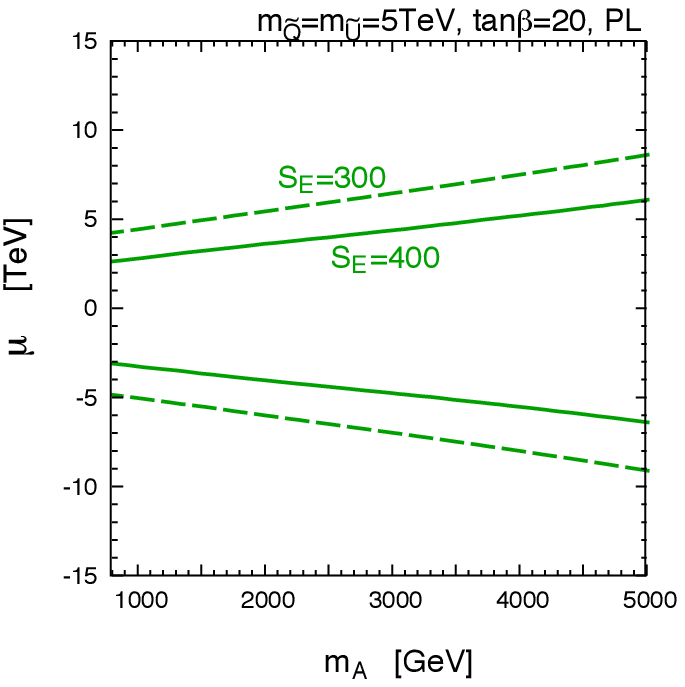} \\
\end{center}
\caption{Contours of $S_E=300$ and $400$ on $m_A$ vs.\ $\mu$ plane for
  the PS and PL solutions of $A_t$.  Here,
  $m_{\tilde{Q}}=m_{\tilde{U}}=M_3=5\TeV$,
  $m_{\tilde{D}}=m_{\tilde{L}}=m_{\tilde{E}}=\max(m_{\tilde{U}},
  |\mu|)$, and $\tan\beta=20$.}
\label{fig:se}
\end{figure}

In Fig.~\ref{fig:se} we show the contours of constant $S_E$ for the PS
and PL solutions on $m_A$ vs.\ $\mu$ plane.  Here,
$m_{\tilde{Q}}=m_{\tilde{U}}=M_3=5\TeV$, while sfermion masses,
$m_{\tilde{D}}$, $m_{\tilde{L}}$ and $m_{\tilde{E}}$, are taken to be
${\rm max} (m_{\tilde{U}}, \mu)$ (see Sec.~\ref{sec:width}).  In each
plot, the region between the solid (dashed) green lines
satisfies $S_E>400$ ($300$).  For the PS solution, these lines appear
only for $|\mu| > 10\TeV$.  On the other hand, the upper bound on
$|\mu|$ is comparable to the stop masses for the PL solution.

The tree-level potential Eq.~\eqref{eq:stophiggs} is used for our 
numerical analysis. 
Radiative corrections may change the scalar potential particularly 
around the SM-like vacuum.  
In order to study their effects, we also estimated the decay rate 
of the SM-like vacuum by including the top-stop and bottom-sbottom 
one-loop correction to the Higgs potential.\footnote
{ A better treatment would be to introduce full one-loop radiative
  corrections to the potential involving stop, sbottom and the Higgs
  boson so that the potential is stable against the renormalization
  scale at least at the one-loop level. }
We found that the Euclidean action tends to increase by about 10 --
20\% for $S_E \sim 400$ from that of the tree-level potential.  In
order to see the sensitivity of $S_E$ to $\mu$, we show the contours
of $S_E = 300$ in the same plot.

When both $\mu$ and $\tan\beta$ are large, we might have to consider 
other CCB vacua in the sbottom-Higgs direction, which is driven by the
tri-linear coupling of $y_b(A_b h_d - \mu h_u)\tilde b_L \tilde
b_R/\sqrt{2}$ (where $\tilde b_L$ and $\tilde b_R$ are left- and
right-handed sbottoms, respectively) (cf.~Ref.~\cite{Altmannshofer:2012ks}).  
If both $\mu$ and $A_b$ are
large, the bottom Yukawa coupling can be enhanced not only by
$\tan\beta$ but also by $(1+\Delta_b)^{-1}$ (see Eq.~\eqref{yb(MSUSY)}).  
However, when the squark masses are universal, such a parameter region 
is already excluded by the other constraints discussed in this section.  
Therefore we do not consider the constraint coming from the CCB 
vacuum involving the sbottom sector in this paper. 


\subsubsection{Bottom Yukawa coupling}

When $\tan\beta$ is very large, the bottom Yukawa coupling is sizable.
We can impose an upper bound on $\tan\beta$ by requiring perturbativity
of the Yukawa coupling constants up to, for instance, the GUT scale.
In the MSSM, the bottom Yukawa coupling constant $y_b$ is proportional
to $(1+\Delta_b)^{-1}$, and hence, $y_b$ is enhanced when $\Delta_b$
is negative (see Eq.~\eqref{m_b}).  Consequently, the bound on
$\tan\beta$ is more severe when $\Delta_b<0$.

In our numerical analysis, we estimate the bottom Yukawa coupling
constant using the following relation:\footnote
{In discussing the perturbative bound, we neglect holomorphic
  corrections to the Yukawa coupling constant, which are
  orders-of-magnitude smaller than the tree-level value of the
  Yukawa coupling constant for $\tan\beta\gg 1$.}
\begin{align}
  y_b(M_{\rm SUSY}) \simeq
  \frac{\sqrt{2}m_b(M_{\rm SUSY})}{v\cos\beta(1+\Delta_b)}.
  \label{yb(MSUSY)}
\end{align}
Then, we follow the evolutions of coupling constants by solving the
renormalization group equations at the one-loop level and impose a
condition that the bottom Yukawa coupling is perturbative up to the
grand unified theory (GUT) scale $M_{\rm GUT}$.  Numerically, we 
require  
\begin{align}
  \left|y_b(M_{\rm GUT})\right|<1,
\end{align}
where we take $M_{\rm GUT}=2\times10^{16}\GeV$.
This constraint excludes a large $\tan\beta$ region especially when
$\Delta_b < 0$.  Notice that $\Delta_b$ is a non-decoupling parameter,
and hence, this constraint is important even in the limit of heavy
superparticles.

\section{Higgs Partial Decay Widths}
\label{sec:width}
\setcounter{equation}{0}


In this section,we discuss the partial decay width of the lightest Higgs boson
$h$.  We define the ratio of the partial decay width of the lightest
Higgs boson to that of the SM prediction:
\begin{align}
  R_{F} \equiv
  \frac{\Gamma (h\rightarrow F)}
  {\Gamma^{\rm (SM)} (h\rightarrow F)},
\end{align}
where $F$ denotes a specific final state.  In the following, we show
how much $R_F$ can deviate from the SM prediction ($R_F=1$) for
various final states. As we have mentioned, {\tt FeynHiggs} 2.10.2 is
used to calculate the partial decay widths of the Higgs boson, in
which full one-loop contributions are taken into account for the
fermionic final states.  In the package, a resummation of the
$\Delta_b$ corrections is also included in calculating the partial
decay width for $h\rightarrow \bar{b}b$ \cite{Williams:2011bu}.

In our numerical calculation, we adopt (approximate) GUT relation 
among the $SU(3)_C$, $SU(2)_L$, and $U(1)_Y$ gaugino masses:
\begin{align}
  M_3 (M_{\rm SUSY}) = 3M_2 (M_{\rm SUSY}) = 6M_1 (M_{\rm SUSY}).
  \label{GUTrelation}
\end{align}
All the phases in the MSSM parameters are assumed to be negligible, 
and we adopt the convention of $M_3>0$.  For simplicity, we also assume 
that the sfermion masses are universal with respect to the generation indices.

\begin{figure}[t]
\begin{center}
\includegraphics[scale=1]{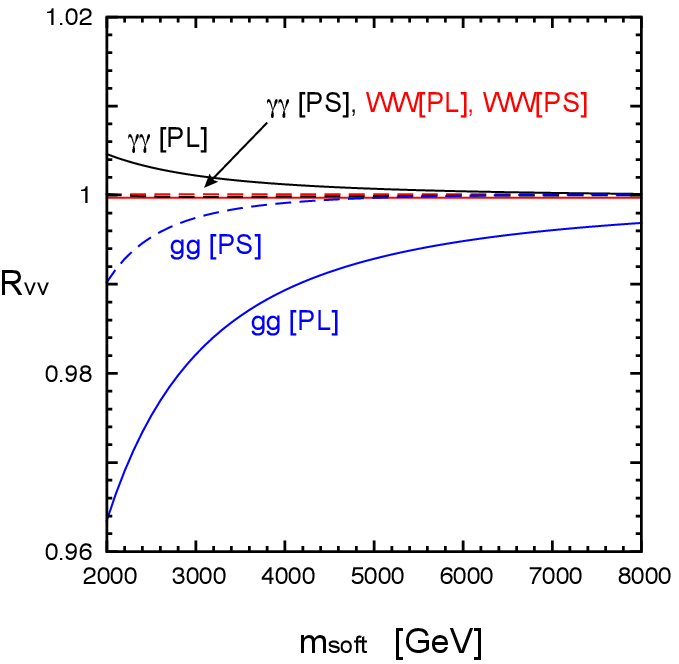}
\includegraphics[scale=1]{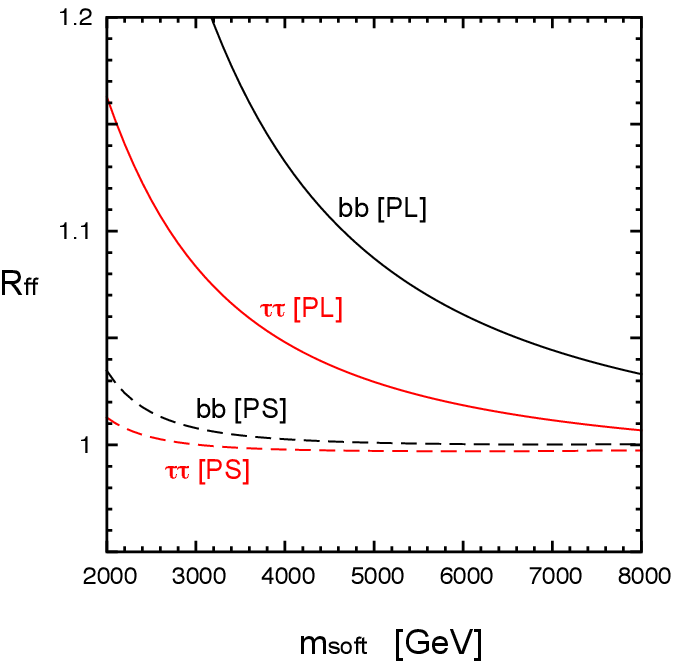}
\end{center}
\caption{$R_{VV}$ (left) and $R_{\bar{f}f}$ (right) are shown as
  functions of $m_{\rm soft}$; here all the fermion masses, $M_3$, and
  $|\mu|$ are equal to $m_{\rm soft}$, and $\tan\beta=40$.  
  (The sign of $\mu$ is taken to
  be negative, while Wino and Bino masses are given by using the
  approximate GUT relation \eqref{GUTrelation}.)  In the left (right)
  plot, the black, red, and blue lines correspond to $h\rightarrow 
  \gamma\gamma$ ($\bar{b}b$), $h\rightarrow W^+W^-$
  ($\bar{\tau}\tau$), and $h\rightarrow gg$,
  respectively.  The solid (dashed) lines correspond to the PL (PS)
  solutions of $A_t$.  Note that the constraints in
  Sec.~\ref{sec:constraint} are not taken into account.  }
\label{fig:msoft}
\end{figure}

First, we show the soft mass dependence of $R_{F}$ without taking the
phenomenological constraints into account.  In Fig.~\ref{fig:msoft},
$R_{F}$ are shown for $F=\gamma\gamma$, $W^+W^-$, $gg$, $\bar{b}b$, and
$\bar{\tau}\tau$ as functions of $m_{\rm soft}$,
taking $M_3=m_A
=-\mu=m_{\tilde{Q}}=m_{\tilde{U}}=m_{\tilde{D}}=m_{\tilde{L}}
=m_{\tilde{E}} \equiv m_{\rm soft}$ and $\tan\beta=40$.\footnote
{ We have checked that $R_{W^+W^-} \simeq R_{ZZ}$ holds in the parameter
  region of our study.  }
By choosing negative $\mu$ with $|\mu|\sim m_{\rm soft}$, the
correction to the bottom Yukawa coupling becomes significant.

Remarkably, even though $m_{\rm soft}$ is around several TeV, the
partial decay widths for $F=\bar{b}b$, $\bar{\tau}\tau$ and $gg$
deviate from the SM prediction by more than $\mathcal{O}(1)$\,\% for
the PL solution.  However, SUSY contributions to the other widths are
smaller.  Note that $R_{\bar{b}b}-1$ is as about twice as
$R_{\bar{\tau}\tau}-1$ for the PL solution, which is due to the
difference between $\Delta_b$ and $\Delta_{\tau}$.  We also note here
that the partial decay widths have an appropriate decoupling
behaviour, i.e., $\Gamma(h\rightarrow F)\rightarrow\Gamma^{\rm
  (SM)}(h\rightarrow F)$ as the masses of superparticles and the heavy
Higgs bosons become infinitely large.

\begin{figure}
\begin{center}
\includegraphics[scale=1]{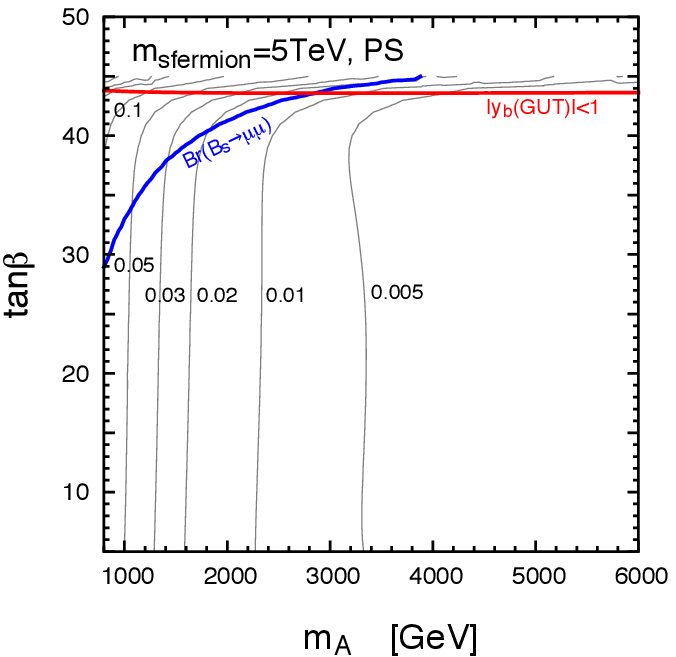} 
\includegraphics[scale=1]{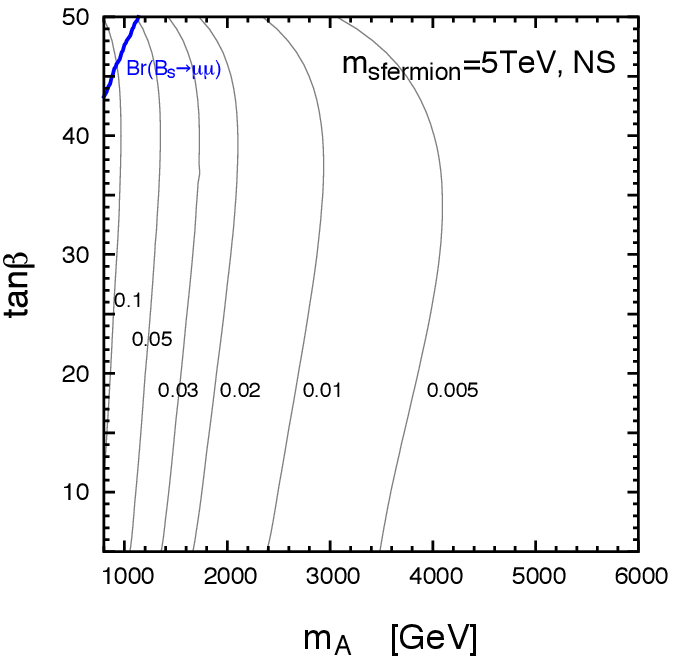} \\
\includegraphics[scale=1]{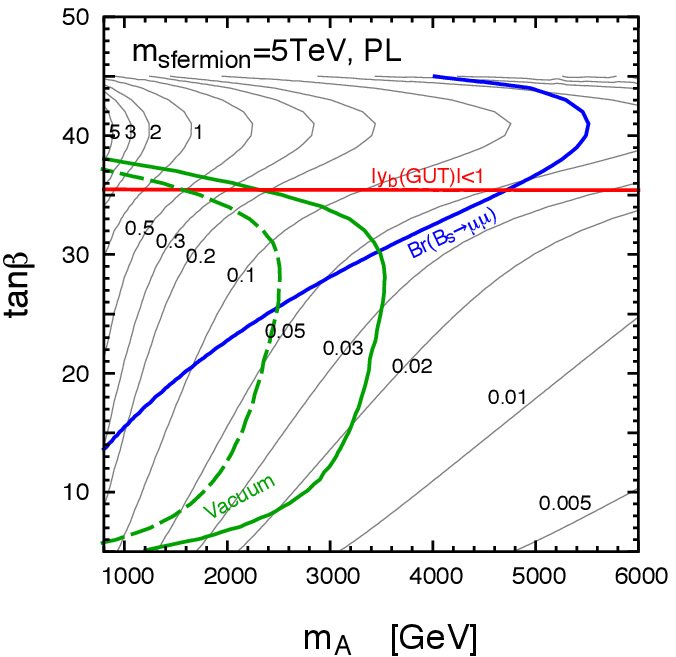} 
\includegraphics[scale=1]{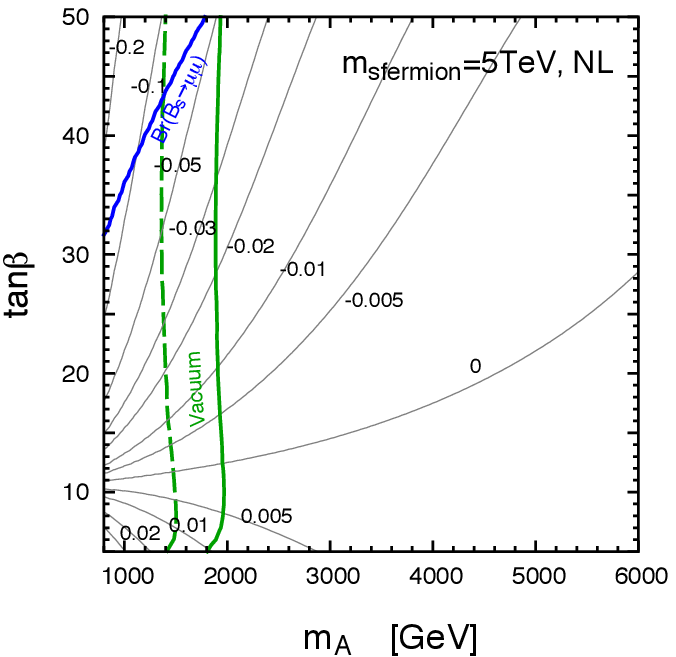}
\end{center}
\caption{Contours of $R_{\bar{b}b} - 1$ are shown for the PS, NS, PL,
  and NL solutions of $A_t$. Here, all the sfermion masses, $M_3$, and
  $|\mu|$ are taken to be $5\TeV$, and the sign of $\mu$ is set to be
  negative.  The left regions of the blue lines are excluded by ${\rm
    Br}(B_s \to \mu^+\mu^-)$, while those of the green solid (dashed)
  lines are constrained by the vacuum stability condition, $S_E > 400$
  (360). The bottom Yukawa coupling becomes non-perturbative below the
  GUT scale in the region above the red line.}
\label{fig:mAtanb}
\end{figure}

Next, let us include the
phenomenological constraints discussed in Sec.~\ref{sec:constraint}.
In Fig.~\ref{fig:mAtanb}, contours of $R_{\bar{b}b}-1$ are shown on
$m_A$ vs.\ $\tan\beta$ plane.  Here, all the sfermion masses, $M_3$,
and $|\mu|$ are set to be $5\TeV$.  In the cases of the
PS and PL solutions, the contours end at around $\tan\beta\sim 45$.
This is because, when $\tan\beta$ is large, there is no
solution for $A_t$ to satisfy $m_h = 125.7\GeV$.  The bottom-sbottom
loop contribution interferes destructively with the top-stop 
contribution, and thus, larger (smaller) $A_t$ is required for the PS
(PL) solution. Then, $A_t^{({\rm PS})}$ and $A_t^{({\rm PL})}$ merge
into a single solution at certain $\tan\beta$, which is the value
where the top-stop contributions cannot raise the Higgs mass
anymore for fixed $m_{\rm soft}$.  Note that, $\Delta_b$ is negative
when $\mu < 0$ and $A_t>0$.  In such a case, $y_b$ is significantly
larger than the tree-level value, and thus, the bottom-sbottom loop
contributions are enhanced.

In Fig.~\ref{fig:mAtanb}, the constraints from ${\rm Br}(B_s \to
\mu^+\mu^-)$, the vacuum stability, and the perturbativity of the
bottom Yukawa couplings are shown.  Wide parameter region is
excluded when $A_t$ is large, i.e., in the PL and NL panels.  The
constraints from ${\rm Br}(B_s \to \mu^+\mu^-)$ and the vacuum
stability become weaker for large $m_A$, while that from the
perturbativity is not.  For the PL solution, the constraints from 
${\rm Br}(B_s \to \mu^+\mu^-)$ and the vacuum stability change 
drastically as we vary $\tan\beta$ for $\tan\beta \gtrsim 30$, where the
bottom Yukawa coupling is much larger than the tree-level value.

\begin{table}[t]
  \begin{center}
    \begin{tabular}{l|cc|cccc|cc}
      \hline\hline
      & 
      \multicolumn{2}{c|}{LHC\,\cite{Dawson:2013bba}} & 
      \multicolumn{4}{c|}{ILC\,\cite{Peskin:2013xra}} &
      \multicolumn{2}{c}{TLEP\,\cite{Dawson:2013bba}}
      \\
      \hline
      $\sqrt{s}$ [GeV] & 
      1400 & 1400 & 
      250 & 500 & \multicolumn{2}{c|}{1000} &
      240 & 350
      \\
      \hline
      $\int\!dt\,\mathcal{L}$ [$\invfb$] & 
      300 & 3000 & 
      250 & 500 & 1000 & 2500 &
      10000 & +2600
      \\
      \hline
      $\gamma\gamma$ & 
      10 -- 14 & 4 -- 10 & 
      38 & 17 & 5.8 & 3.8 & 
      3.4 & 3.0
      \\
      $gg$ & 
      12 -- 16 & 6 -- 10 & 
      12 & 4.0 & 1.6 & 1.2 &
      2.2 & 1.6
      \\
      $\bar{b}b$ & 
      20 -- 26 & 8 -- 14 & 
      9.4 & 1.9 & 0.78 & 0.64 &
      1.8 & 0.84
      \\
      $\bar{\tau}\tau$ & 
      12 -- 16 & 4 -- 10 & 
      10 & 3.8 & 1.6 & 1.3 &
      1.9 & 1.1
      \\
      \hline\hline
    \end{tabular}
    \caption{Expected accuracies of the determinations of the partial
      decay widths of the Higgs boson in units of percents
      \cite{Dawson:2013bba,Peskin:2013xra}.  The accuracies of the
      widths are assumed to be twice the accuracies of the
      determinations of couplings.}
    \label{table:experiments}
  \end{center}
\end{table}

Even if the masses of superparticles are relatively large (i.e.,
$5\TeV$), $R_{\bar{b}b}-1$ can be as large as $\mathcal{O}(1)$\,\%.
Such a large deviation may be within the reach of future collider
experiments.  Expected accuracies at the future experiments have been
discussed (see Table 1-16 of Ref.~\cite{Dawson:2013bba} and
Ref.~\cite{Peskin:2013xra}); the numbers are summarized in
Table~\ref{table:experiments}.  The accuracies of $\delta
\Gamma(h\rightarrow \bar{b}b)=0.64\%$, $\delta \Gamma(h\rightarrow
\bar{\tau}\tau)=1.1\%$, and $\delta \Gamma(h\rightarrow gg)=1.2\%$ are
claimed to be achievable at $e^+e^-$ colliders ultimately.  Therefore,
it is found that, even if the superparticles are kinematically
unaccessible at the LHC, we may observe the MSSM signal by studying 
the partial decay widths of $h$ in detail.

For the PL solution, it is also found that the partial width of
$h\rightarrow \bar{b}b$ can deviate from the SM prediction by about
2\% even for $m_A=6\TeV$.  On the other hand, we checked that
$R_{\bar{\tau}\tau}$ is smaller by about 1\% than $R_{\bar{b}b}$ at
the same parameter point, because the non-holomorphic correction
$\vert \Delta_b\vert $ is bigger than $\vert \Delta_{\tau}\vert$ at
this model point. When $\tan\beta$ is smaller, the difference between
$R_{\bar{b}b}$ and $R_{\bar{\tau}\tau}$ decreases, since $\Delta_b$ is
approximately proportional to $\tan\beta$; in such a case both
$R_{\bar{b}b}$ and $R_{\bar{\tau}\tau}$ are well approximated by
$(\sin\alpha/\cos\beta)^2$.

\begin{figure}
\begin{center}
\includegraphics[scale=1.]{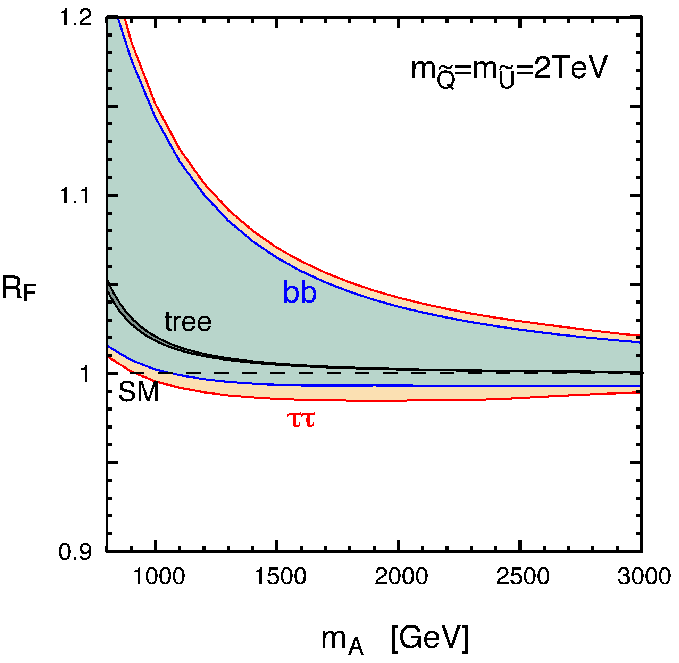} \hspace{5mm}
\includegraphics[scale=1.]{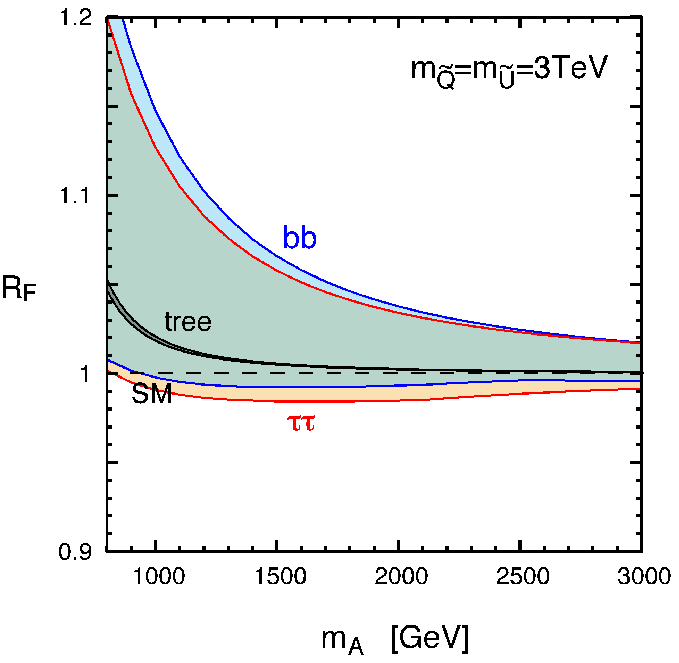} \\ \vspace{5mm}
\includegraphics[scale=1.]{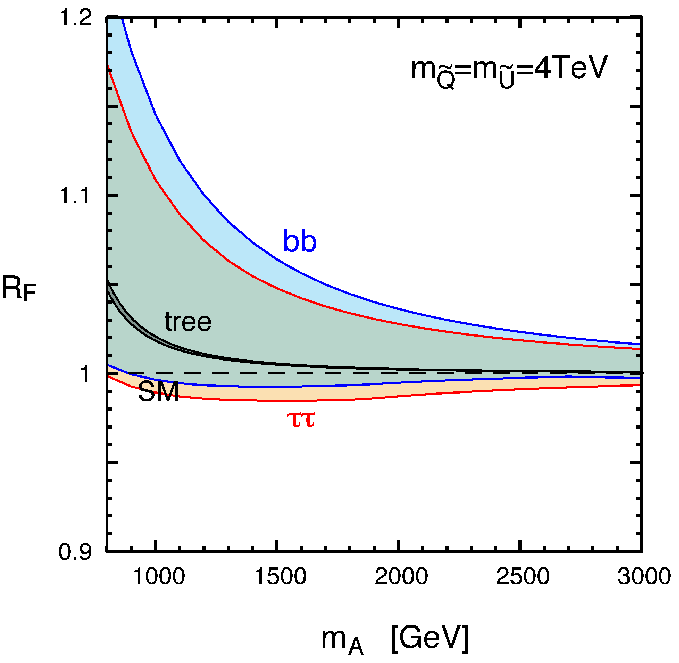} \hspace{5mm}
\includegraphics[scale=1.]{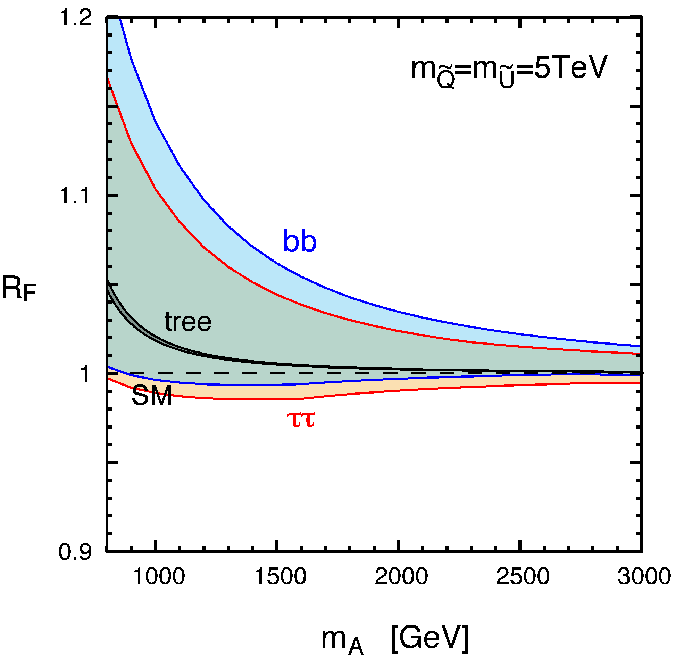}
\end{center}
\caption{The ranges of $R_{\bar{b}b}$ (blue) and $R_{\bar{\tau}\tau}$
  (red) with small $|A_t|$ solutions (i.e, the PS and NS solutions for
  $A_t$).  We take $m_{\tilde{Q}}=m_{\tilde{U}}=M_3=2$, 3, 4 and
  $5\TeV$.  Other sfermion masses are set to be $\max(m_{\tilde{Q}},
  |\mu|)$. The partial widths are constrained by ${\rm Br}(B_s
  \to \mu^+\mu^-)$, the vacuum stability condition ($S_E > 400$), 
  and the perturbativity of the bottom Yukawa coupling 
  ($|y_b({\rm GUT})| < 1$). The black region corresponds to the 
  tree-level prediction for $\tan\beta=5$--$50$.
  The black dashed line is the SM prediction. }
\label{fig:RbbtauS}
\end{figure}

\begin{figure}
\begin{center}
\includegraphics[scale=1.]{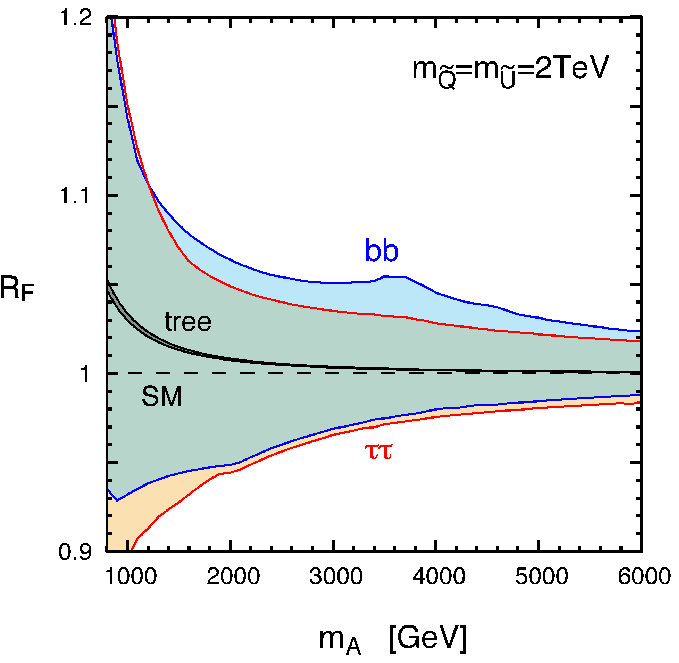} \hspace{5mm}
\includegraphics[scale=1.]{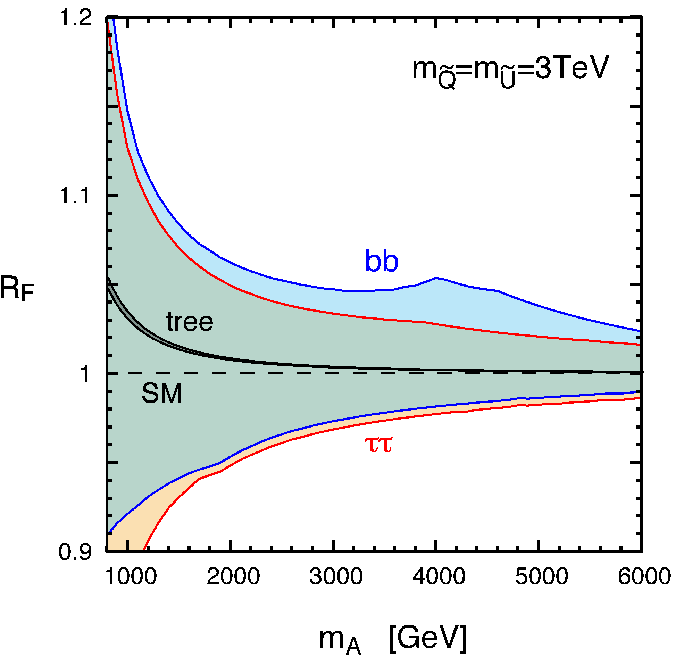} \\ \vspace{5mm}
\includegraphics[scale=1.]{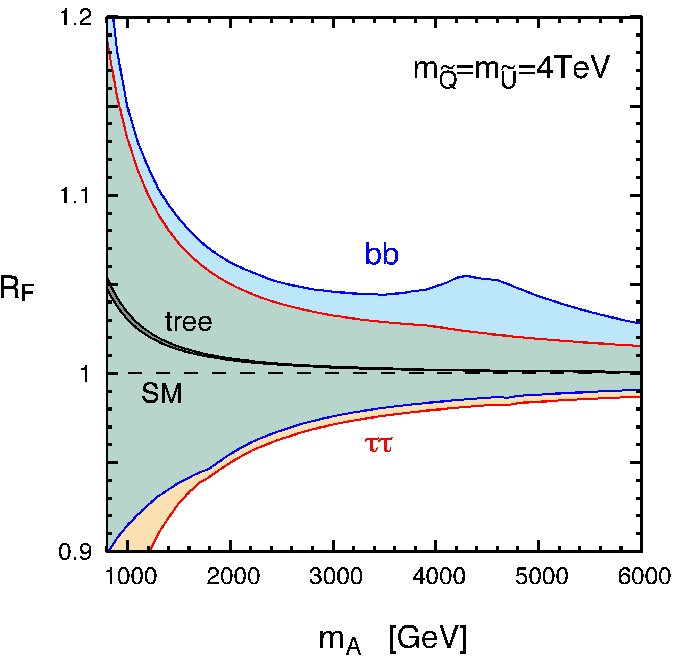} \hspace{5mm}
\includegraphics[scale=1.]{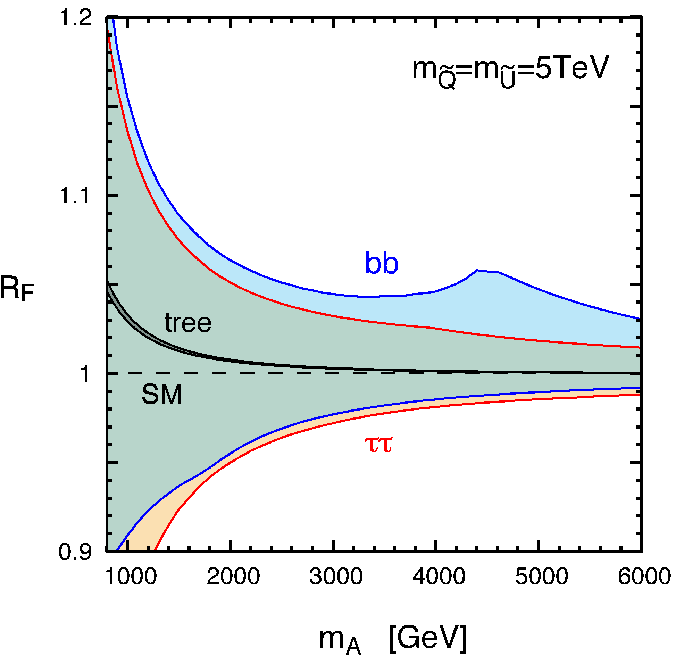}
\end{center}
\caption{Same as Fig.~\ref{fig:RbbtauS}, but with all the 
  solutions of $A_t$.}
\label{fig:Rbbtau}
\end{figure}

Let us see how much $R_{\bar{b}b}$ and $R_{\bar{\tau}\tau}$ can
change in the parameter space consistent with the phenomenological 
constraints. We have performed the scan in the following parameter 
space of the MSSM:
\begin{itemize}
\item $m_{\tilde{Q}}=m_{\tilde{U}}=M_3=2$, $3$, $4$, and $5\TeV$,
\item $m_{\tilde{D}}=m_{\tilde{L}}=m_{\tilde{E}}=\max(m_{\tilde{U}},
  |\mu|)$,
\item $A_t=A_t^{({\rm NS})}$, $A_t^{({\rm NL})}$, 
  $A_t^{({\rm PS})}$, $A_t^{({\rm PL})}$,
\item $0.8\TeV \leq m_A \leq 6\TeV$,
\item $-5 \leq \mu/m_{\tilde{U}} \leq -0.5$, or $0.5 \leq \mu/m_{\tilde{U}} \leq 5$,
\item $5 \leq \tan\beta \leq 50$.
\end{itemize}
Here, $A_t$ is determined to satisfy $m_h=125.7\GeV$.  
The other tri-linear couplings (e.g., $A_b$ and $A_\tau$) are
assumed to be equal to $A_t$; we checked that our numerical results
are insensitive to this assumption.  We take
$m_{\tilde{Q}}=m_{\tilde{U}}=M_3$, while sfermion masses other than
$m_{\tilde{Q}}$ and $m_{\tilde{U}}$ are set to be equal to
$\max(m_{\tilde{U}}, |\mu|)$.  We have checked that $R_{\bar{b}b}$ and
$R_{\bar{\tau}\tau}$ are almost insensitive to these scalar masses
unless they are very small.  However, when $|\mu|$ is much larger than
$m_{\tilde{D}}$, $m_{\tilde{L}}$, and $m_{\tilde{E}}$, bottom and stau
mixings become sizable which causes additional complexity to our
parameter scan, and/or lighter stau become tachyonic.  The values of
$m_{\tilde{D}}$, $m_{\tilde{L}}$, and $m_{\tilde{E}}$ are chosen to
avoid this problem.

If $|\mu|$ would become much larger than squark masses
(e.g., $|\mu|/m_{\tilde{U}} \gg 5$), the Higgs partial widths could
deviate from the SM prediction sizably, since the Higgs mixing angle
would be enhanced by radiative corrections.  However, the
vacuum stability condition excludes significant amount of the
parameter region with large $|\mu|$, as we have shown in Fig.~\ref{fig:se}.  
In our scan, we set $|\mu|/m_{\tilde{U}} \leq 5$; this upper bound is 
large enough to cover the whole region allowed by the vacuum stability.

In Figs.~\ref{fig:RbbtauS} and \ref{fig:Rbbtau}, we show the minimal
and maximal values of $R_{\bar{b}b}$ and $R_{\bar{\tau}\tau}$ as
functions of $m_A$ and $m_{\rm soft}$, taking into account the
phenomenological constraints discussed in the previous
section. Fig.~\ref{fig:RbbtauS} corresponds to the small $A_t$
solutions (NS and PS). The largest and smallest values of
$R_{\bar{b}b}$ and $R_{\bar{\tau}\tau}$ are achieved when $|\mu|$ and
$\tan\beta$ are large and are marginally allowed by the
phenomenological constraints. The deviations mainly come from
radiative corrections to the Higgs mixing angle and $\Delta_b$. In
particular, the maximal value of $R_{\bar{b}b}$ is larger than that of
$R_{\bar{\tau}\tau}$ for large $m_{\tilde Q}=m_{\tilde U}$. As
discussed earlier, $\Delta_b$ becomes negative and sizable with
$\mu<0$, which results in a significant enhancement of $R_{\bar{b}b}$.

Fig.~\ref{fig:Rbbtau} shows the minimal and maximal values of
$R_{\bar{b}b}$ and $R_{\bar{\tau}\tau}$, taking all the solutions of
NS, PS, NL, and PL into consideration. They change significantly 
compared to Fig.~\ref{fig:RbbtauS}. The partial widths become extremum 
when $A_t$ takes the NL or PL solution except the maximal value 
for small $m_A$.  For $m_A \lesssim 1\TeV$, the NS or PS solutions 
give the maximal value, since the
phenomenological constraints, especially the vacuum stability
condition, are too severe for the NL and PL solutions (see
Fig.~\ref{fig:mAtanb}). As in the case of Fig.~\ref{fig:RbbtauS}, the
maximal value of $R_{\bar{b}b}$ is much larger than that of
$R_{\bar{\tau}\tau}$ because $|\Delta_b| \gg |\Delta_\tau|$.
The maximal value of $R_{\bar{b}b}$ has a non-trivial bump-like structure. 
When $m_A$ is small, medium (i.e., around the peak of the bump), and large, 
the partial decay width is bounded by the vacuum stability, flavor constraint, 
and the perturbativity of $y_b$, respectively. 

The measurement of the Higgs couplings may provide an evidence of BSM,
because the values of $R_{\bar{b}b}$ and $R_{\bar{\tau}\tau}$ may show
significant deviation from the SM prediction.  In future experiments,
the partial decay widths of $h \rightarrow \bar{b}b$ and $h \rightarrow
\bar{\tau}\tau$ may be measured at the $\lesssim1\%$ level (see Table
\ref{table:experiments}).  In the case of the PS and NS solutions,
$R_{\bar{b}b}-1$ can be as large as 2\% (3\%) if $m_A$ is $2.7\TeV$
($2.2\TeV$) for $m_{\rm soft}=2\TeV$, and $2.6\TeV$ ($2.1\TeV$) for
$m_{\rm soft}=5\TeV$. On the other hand, $R_{\bar{\tau}\tau}-1$
becomes larger than 2\% (3\%) when $m_A$ is smaller than $3.0\TeV$
($2.4\TeV$) for $m_{\rm soft}=2\TeV$, and $2.1\TeV$ ($1.8\TeV$) for
$m_{\rm soft}=5\TeV$.  Including the PL and NL solutions, deviations 
of $2$ -- $3$\% level are achieved with larger value of 
$m_A$. For example, with $m_{\rm soft}=5\TeV$, $R_{\bar{b}b}-1=3\%$ 
can be achieved with $m_A=6.0\TeV$.

\begin{figure}[t]
\begin{center}
\includegraphics[scale=1]{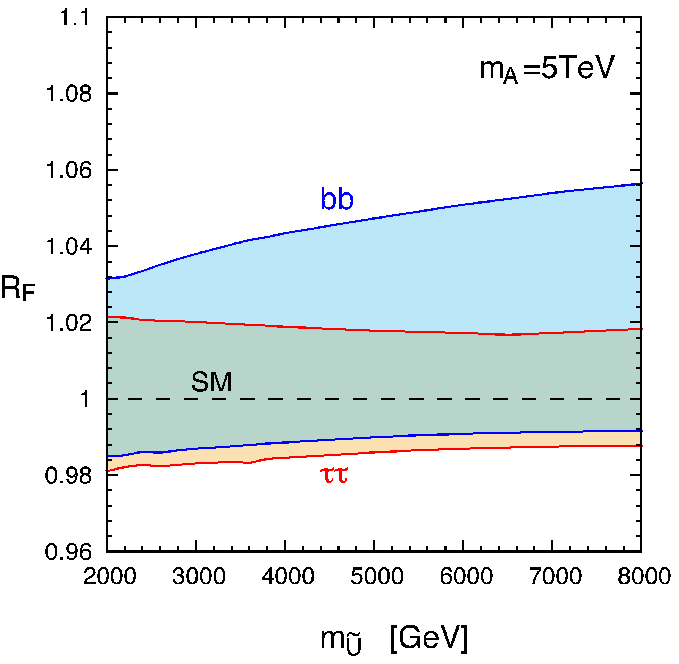}
\includegraphics[scale=1]{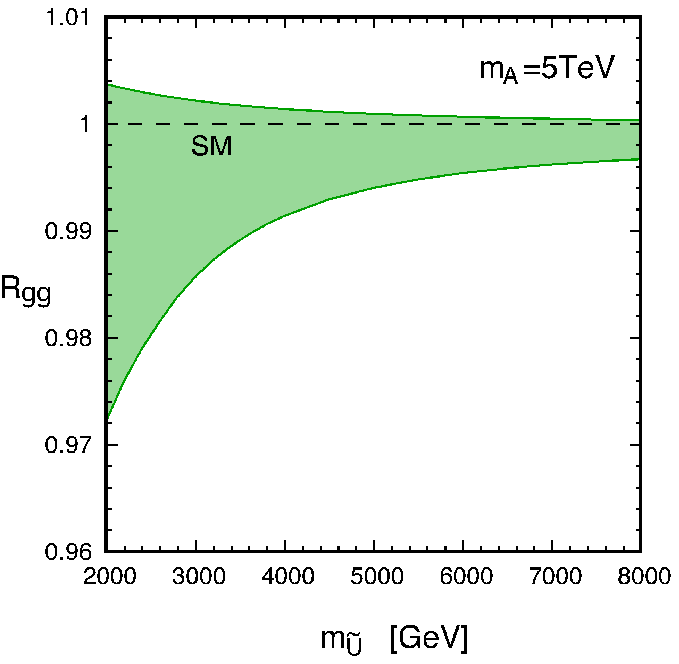}
\end{center}
\caption{The ranges of $R_{\bar{f}f}$ (left) and $R_{gg}$ (right) are
  shown as functions of $m_{\tilde{U}}$. The parameters are scanned in
  the same manner as Fig.~\ref{fig:Rbbtau}, but $m_A$ is fixed to be
  $5\TeV$ while $m_{\tilde{U}}$ is varied.  Here, all the solutions of
  $A_t$ are included in the scan with imposing the phenomenological
  constraints discussed in Sec.~\ref{sec:constraint}. The black dashed
  line is the SM prediction.}
\label{fig:msoft2}
\end{figure}

In Fig.~\ref{fig:msoft2}, we show the sfermion mass dependence of
$R_{F}$ with taking the phenomenological constraints into account.
The parameter scan is performed in the same setup as
Fig.~\ref{fig:Rbbtau}, but $m_{\tilde{U}}$ is varied. Here, $m_A$ 
is fixed to be $5\TeV$. In the left plot, the minimal and maximal
values of $R_{\bar{b}b}$ and $R_{\bar{\tau}\tau}$ are shown. They are
achieved by the PL or NL solution. Since $m_A$ is fixed, the
decoupling behaviour is not observed. Rather, the non-decoupling
contribution to $\Delta_b$ enhances $R_{\bar{b}b}$ as $m_{\tilde{U}}$
increases, since larger $|A_t|$ is required to satisfy $m_h=125.7\GeV$
for the solutions. On the other hand, it is found that radiative
corrections to the Higgs mixing angle do not change so much even if
$m_{\tilde{U}}$ increases.

In the right panel of Fig.~\ref{fig:msoft2}, the minimal and maximal
values of $R_{gg}$ are displayed. The maximal value occurs for the PS
or NS solution, while the minimal value is achieved by the PL or NL
solution.  We can see the decoupling behavior, i.e., $R_{gg}$
approaches to unity as $m_{\tilde{U}}$ increases. The partial decay
width of $h\rightarrow gg$ can deviate from the SM prediction by about
3\% for $m_{\tilde{U}}=2\TeV$, while it decreases rapidly and becomes
1\% for $m_{\tilde{U}}=3.7\TeV$.  We have also checked that these
values do not change so much for $m_A=8\TeV$. (However, they change
significantly if $m_A$ is smaller, since the phenomenological
constraints exclude the parameter space severely.)  
According to Table \ref{table:experiments}, the partial width of 
$h\rightarrow gg$ is expected to be measured at the $1.2\%$ accuracy 
in future experiments. Thus, as far as superparticles are relatively light,
we may observe a signal of the MSSM in the measurements of this partial 
width even if the heavier Higgses are out of the reach of the LHC.

\begin{figure}[t]
\begin{center}
\includegraphics[scale=0.4]{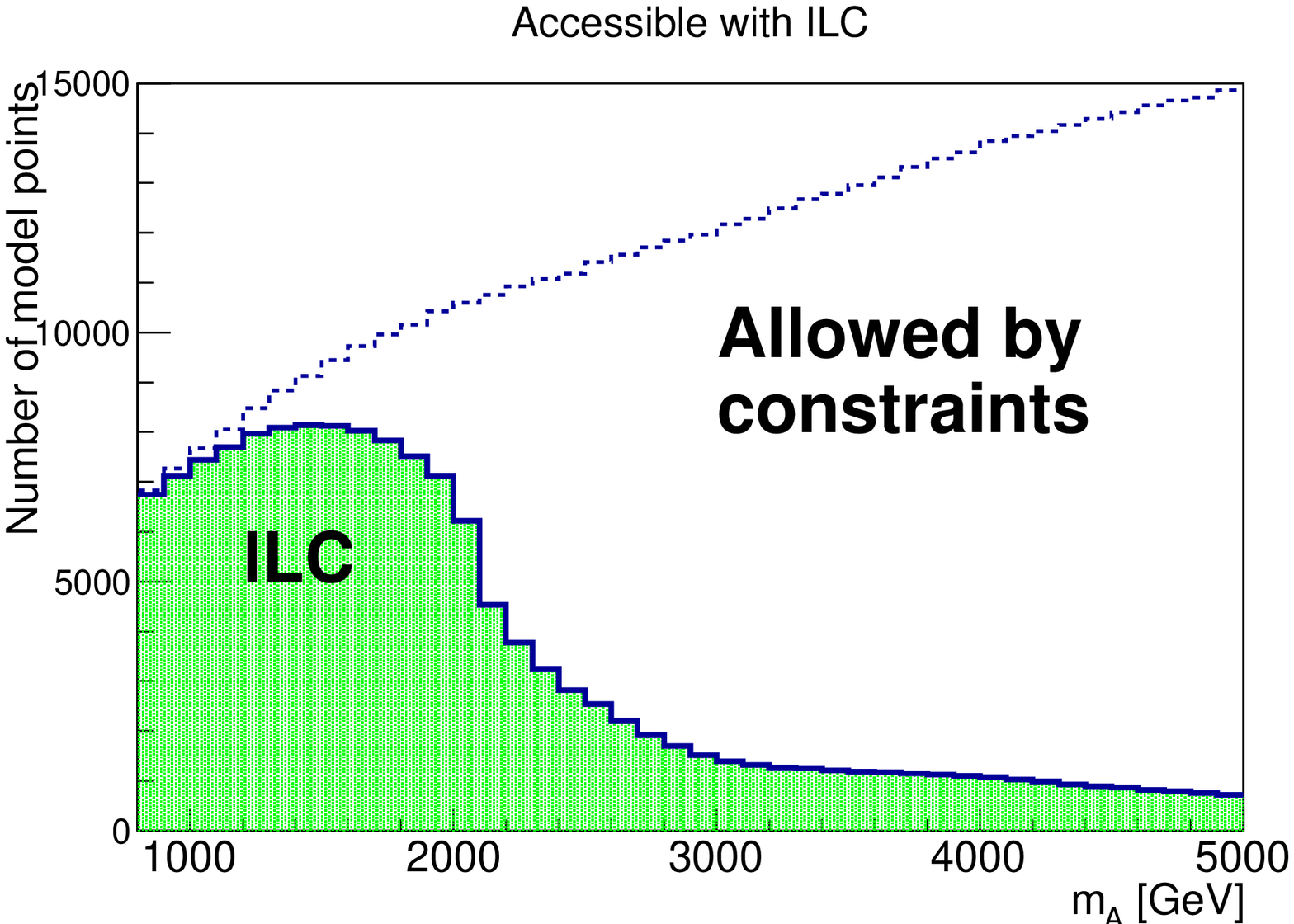} 
\includegraphics[scale=0.4]{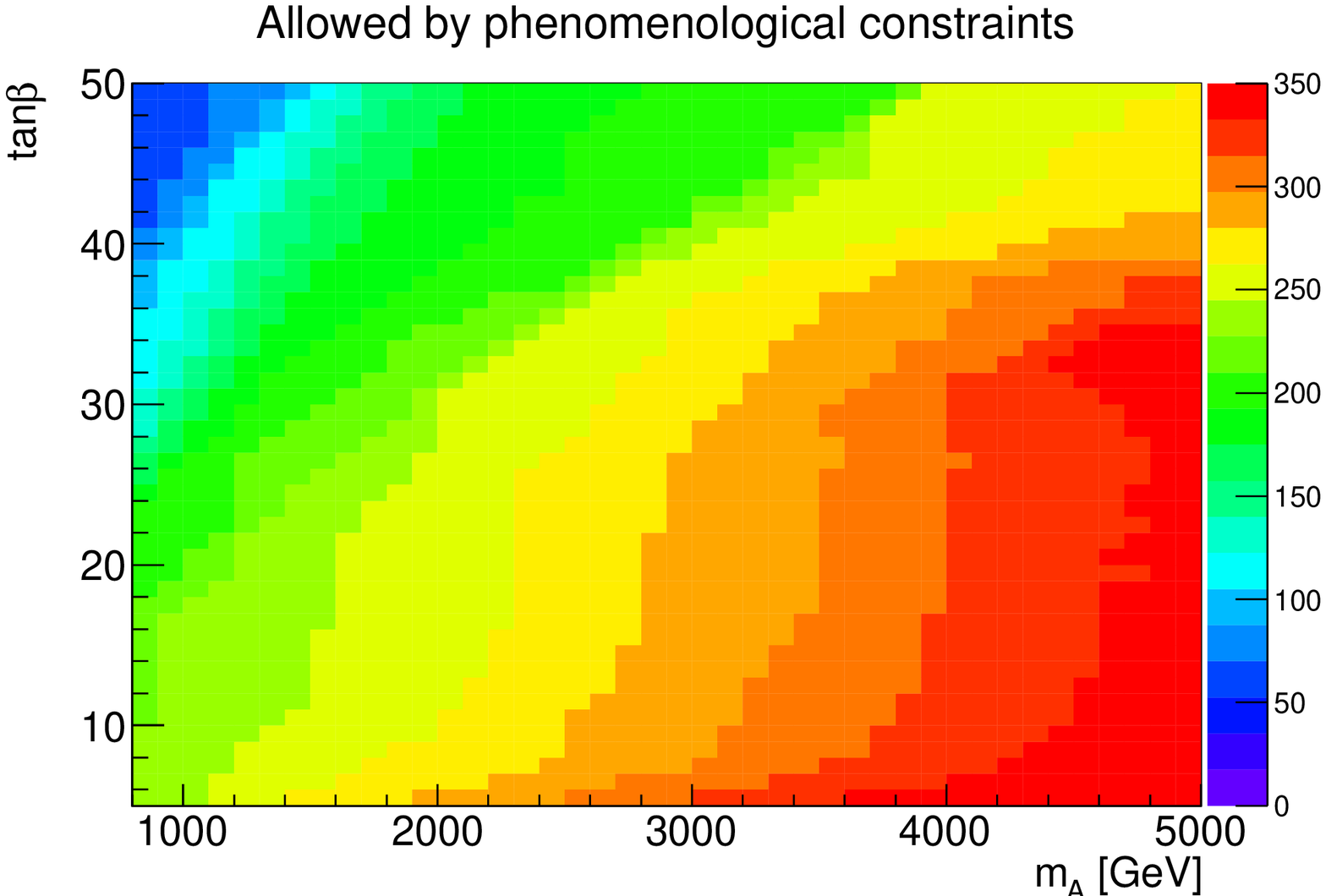} \\
\includegraphics[scale=0.4]{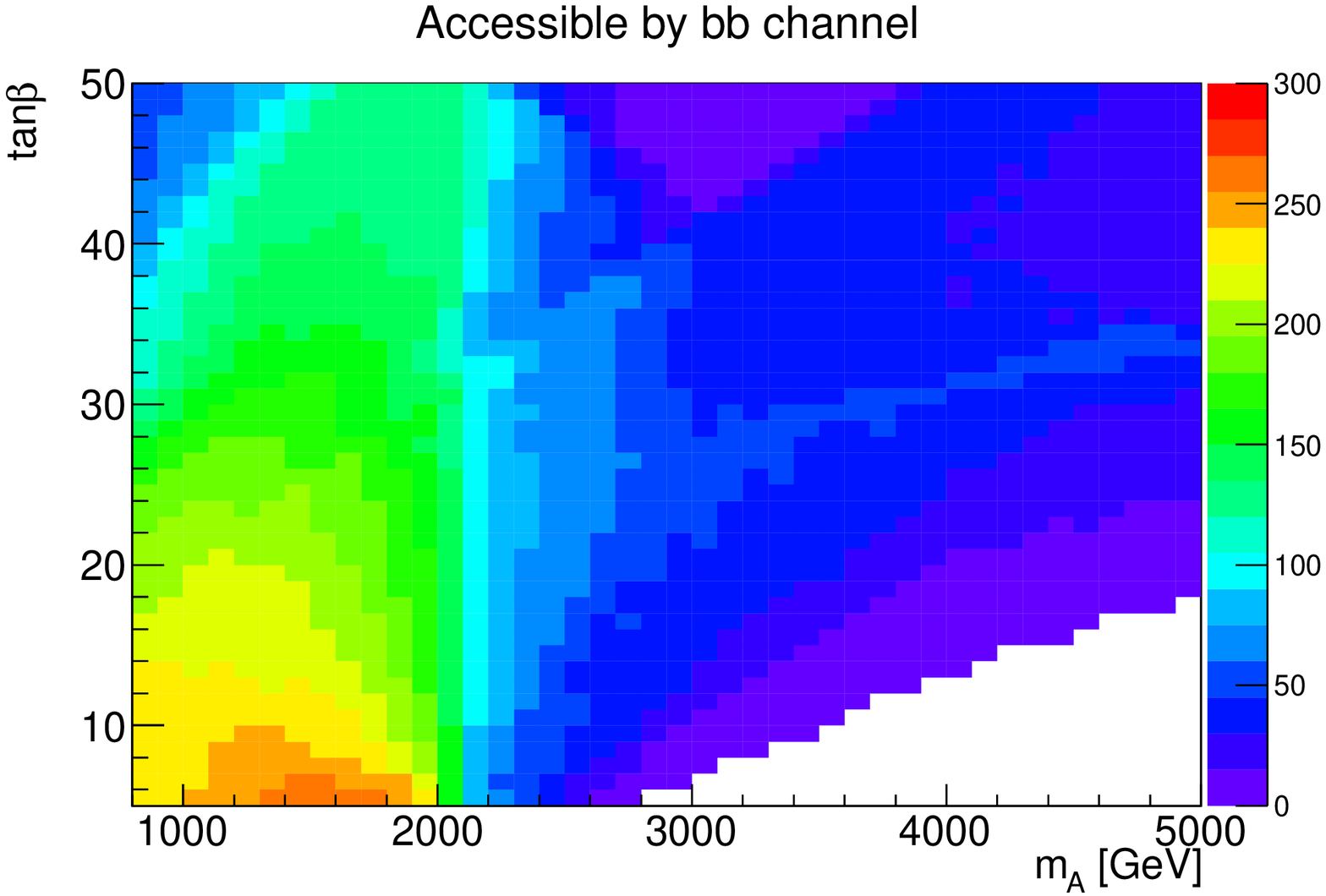} 
\includegraphics[scale=0.4]{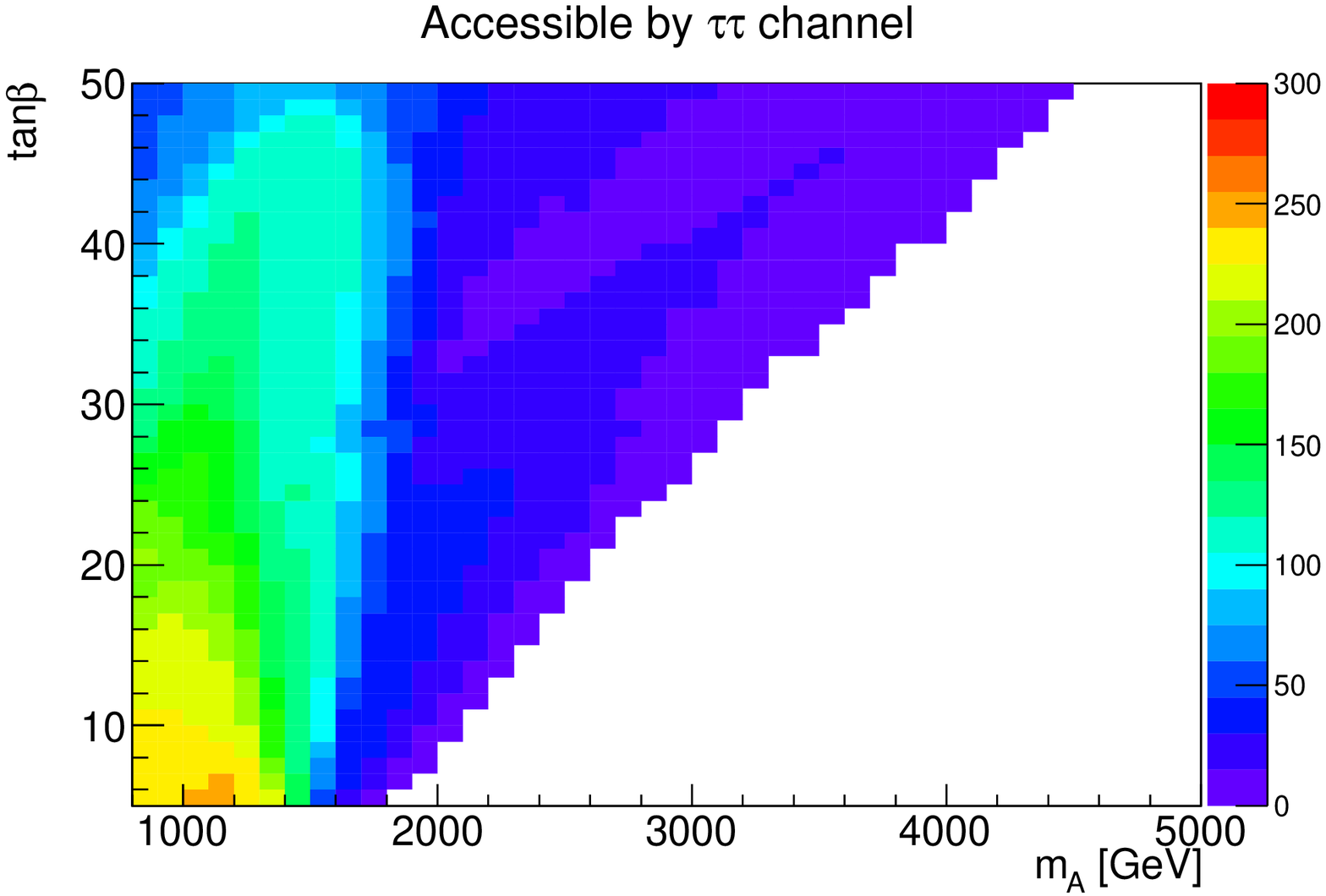}
\end{center}
\caption{Upper-left: The number of model points accessible 
  with ILC by at least one decay mode of $h$ as a function 
  of $m_A$ (green histogram), as well as that of model points 
  allowed by the phenomenological constraints (dotted histogram).  
  Upper-right: The number of model points allowed by the phenomenological 
  constraints on $m_A$ vs.\ $\tan\beta$ plane.  Lower-left: The number 
  of model points accessible with ILC by $h\rightarrow\bar{b}b$.  
  Lower-right: The number of model points accessible with ILC by 
  $h\rightarrow\bar{\tau}\tau$.
}
\label{fig:scan}
\end{figure}

Finally, we show how large fraction of the parameter space can be
covered by future $e^+e^-$ colliders.  For this purpose, we define the
$\delta\chi_F^2$ variable as
\begin{align}
  \delta \chi_F^2 = 
  \frac{\left[\Gamma(h\rightarrow F)-\Gamma^{\rm (SM)}(h\rightarrow F)\right]^2}
  {[\delta\Gamma(h\rightarrow F)]^2},
\end{align}
where $\delta\Gamma(h\rightarrow F)$ is the expected accuracies of the
determinations of the Higgs partial decay widths at ILC with
$\sqrt{s}=1\TeV$ and $\int\!dt\,\mathcal{L}=2500\invfb$ (see Table
\ref{table:experiments}).  Based on this quantity, we define the
parameter region which is accessible with ILC at $\delta \chi_F^2 
\geq 4$.  We perform a parameter scan and study if each model point 
is accessible with ILC and satisfies the phenomenological constraints.
The MSSM parameters are scanned in the ranges of $800\GeV \leq m_A
\leq 5\TeV$ (with the step size of $100\GeV$), $5\leq\tan\beta\leq50$ 
(with the step size of $1$), and $0.5\leq|\mu|/m_{\tilde{U}}\leq2$ 
(with the step size of $0.1$). Thus, for each set of $(m_A, \tan\beta)$, 
384 model points are studied, taking account of positive and negative 
values of $\mu$ as well as all four solutions of $A_t$. In addition, 
we adopt the relations $m_{\tilde{Q}}=M_3=m_{\tilde{U}}$, and 
$m_{\tilde{D}}=m_{\tilde{L}}=m_{\tilde{E}}=\max(m_{\tilde{U}},|\mu|)$. 
Concentrating on the parameter space where the LHC will have a difficulty 
in finding superparticles, $m_{\tilde{U}}$ is taken to be $3$, $4$, 
and $5\TeV$.  In Fig.~\ref{fig:scan}, the ILC coverage of the parameter
space of our scan is displayed.  In the upper-left panel, we show the 
distribution of the number of model points accessible with ILC by any 
of the decay modes of $h$ as a function of $m_A$.  The green histogram 
is a distribution of the number of model points which satisfy $\delta 
\chi_F^2 \geq 4$ and the phenomenological constraints, while the  
dotted one is that satisfying the phenomenological constraints without 
imposing $\delta \chi_F^2 \geq 4$.  We find that the number reduces
drastically at $m_A \sim 2\TeV$.  In the upper-right panel, we show
the number of model points which survive the phenomenological 
constraints on $m_A$ vs.~$\tan\beta$ plane. Here, $\delta \chi_F^2 
\geq 4$ is not imposed.  Then, in the lower panels, we show the numbers 
of model points which can be accessed by the decay modes of 
$h\rightarrow\bar{b}b$ (lower-left) and $h\rightarrow\bar{\tau}\tau$ 
(lower-right) with taking account of the phenomenological constraints.  
They reduce significantly for $m_A\gtrsim 2\TeV$ irrespective of 
$\tan\beta$.  The accessible points for $m_A\gtrsim 2\TeV$ are mostly 
with PL or NL solution.

\section{Conclusions and Discussion}
\label{sec:discussion}
\setcounter{equation}{0}

In this paper, we have studied the partial decay widths of the
lightest Higgs boson in the MSSM.  Taking account of relevant
phenomenological constraints, i.e., Higgs mass, flavor constraints,
vacuum stability, and the perturbativity of coupling constants up to
the GUT scale, we have calculated the expected deviations of the 
partial decay widths from the SM predictions.

The partial decay widths are enhanced if the $\mu$-parameter is 
relatively large. However, such a choice may conflict with some 
of the phenomenological constraints.  
In particular, the vacuum-stability condition imposes a
stringent constraint on the parameter space.  We have found that,
with too large $|\mu|$, there show up CCB vacua where the
down-type Higgs field as well as the up-type Higgs and stop fields 
acquire large VEVs; existence of such CCB vacua was not seriously
considered in the previous studies.  In addition, when $\mu\tan\beta$ 
is large, non-holomorphic correction to the bottom Yukawa
interaction becomes so large that the bottom Yukawa coupling constant
becomes non-perturbative below the GUT scale.  Large value of
$\mu\tan\beta$ may also cause too large flavor-violating decay of
$B$-mesons.  By taking these constraints into account, the maximal and
minimal possible values of the Higgs partial widths are restricted.

We found that the deviations of the partial decay widths from the SM
predictions can be of $\mathcal{O}(1)$\,\% for some of the decay
modes.  In particular, those of $\Gamma(h\rightarrow\bar{b}b)$ and
$\Gamma(h\rightarrow\bar{\tau}\tau)$ may show significant deviations even if
the superparticles are out of the reach of 14TeV LHC.  In addition,
the deviation of $\Gamma(h\rightarrow gg)$ may also be sizable if the
superparticles are relatively light.  We emphasize that, although our
scan is limited to some part of the MSSM parameter space, we have found 
the regions where the deviations from the SM predictions
are within the reach of proposed $e^+e^-$ colliders even if
superparticles would not be observed at the LHC.

\vspace{1em}
\noindent {\it Acknowledgements}: 
The authors would like to thank Yasuhiro Shimizu for the 
collaboration in the early stage of this project.
The authors also acknowledge YITP for their hospitality, 
at which this work was initiated.
This work is supported by JSPS KAKENHI No.~23740172 (M.E.), 
No.~26400239 (T.M.), and No.~26287039 (M.M.N.).
The work is supported by Grant-in-Aid for Scientific research 
from the Ministry of Education, Science, Sports, and Culture (MEXT), 
Japan, No.~23104008 (T.M.) and No.~23104006 (M.M.N.), and also by World 
Premier International Research Center Initiative (WPI Initiative), MEXT, 
Japan.




\begin{thebibliography}{99}

\bibitem{Aad:2012tfa}
  G.~Aad {\it et al.}  [ATLAS Collaboration],
  Phys.\ Lett.\ B {\bf 716} (2012) 1
  [arXiv:1207.7214 [hep-ex]].

\bibitem{Chatrchyan:2012ufa}
  S.~Chatrchyan {\it et al.}  [CMS Collaboration],
  Phys.\ Lett.\ B {\bf 716} (2012) 30
  [arXiv:1207.7235 [hep-ex]].

\bibitem{Dawson:2013bba}
  S.~Dawson, A.~Gritsan, H.~Logan, J.~Qian, C.~Tully, R.~Van Kooten, A.~Ajaib and A.~Anastassov {\it et al.},
  arXiv:1310.8361 [hep-ex].

\bibitem{Okada:1990vk}
  Y.~Okada, M.~Yamaguchi and T.~Yanagida,
  Prog.\ Theor.\ Phys.\  {\bf 85} (1991) 1.

\bibitem{Okada:1990gg} 
  Y.~Okada, M.~Yamaguchi and T.~Yanagida,
  Phys.\ Lett.\ B {\bf 262}, 54 (1991).

\bibitem{Ellis:1990nz}
  J.~R.~Ellis, G.~Ridolfi and F.~Zwirner,
  Phys.\ Lett.\ B {\bf 257} (1991) 83.

\bibitem{Ellis:1991zd} 
  J.~R.~Ellis, G.~Ridolfi and F.~Zwirner,
  Phys.\ Lett.\ B {\bf 262}, 477 (1991).

\bibitem{Haber:1990aw}
  H.~E.~Haber and R.~Hempfling,
  Phys.\ Rev.\ Lett.\  {\bf 66} (1991) 1815.

\bibitem{Skands:2003cj}
  P.~Z.~Skands, B.~C.~Allanach, H.~Baer, C.~Balazs, G.~Belanger, F.~Boudjema, A.~Djouadi and R.~Godbole {\it et al.},
  JHEP {\bf 0407} (2004) 036
  [hep-ph/0311123].

\bibitem{PDG}
  K.~A.~Olive et al.~(Particle Data Group), 
  Chin.\ Phys.\ C, {\bf 38}, 090001 (2014).

\bibitem{Heinemeyer:1998yj} 
  S.~Heinemeyer, W.~Hollik and G.~Weiglein,
  Comput.\ Phys.\ Commun.\  {\bf 124}, 76 (2000)
  [hep-ph/9812320].

\bibitem{Heinemeyer:1998np} 
  S.~Heinemeyer, W.~Hollik and G.~Weiglein,
  Eur.\ Phys.\ J.\ C {\bf 9}, 343 (1999)
  [hep-ph/9812472].

\bibitem{Degrassi:2002fi} 
  G.~Degrassi, S.~Heinemeyer, W.~Hollik, P.~Slavich and G.~Weiglein,
  Eur.\ Phys.\ J.\ C {\bf 28}, 133 (2003)
  [hep-ph/0212020].

\bibitem{Frank:2006yh} 
  M.~Frank, T.~Hahn, S.~Heinemeyer, W.~Hollik, H.~Rzehak and G.~Weiglein,
  JHEP {\bf 0702}, 047 (2007)
  [hep-ph/0611326].

\bibitem{Hahn:2013ria}
  T.~Hahn, S.~Heinemeyer, W.~Hollik, H.~Rzehak and G.~Weiglein,
  Phys.\ Rev.\ Lett.\  {\bf 112} (2014) 14,  141801
  [arXiv:1312.4937 [hep-ph]].

\bibitem{Hall:1993gn} 
  L.~J.~Hall, R.~Rattazzi and U.~Sarid,
  Phys.\ Rev.\ D {\bf 50}, 7048 (1994)
  [hep-ph/9306309, hep-ph/9306309].

\bibitem{Hempfling:1993kv} 
  R.~Hempfling,
  Phys.\ Rev.\ D {\bf 49}, 6168 (1994).

\bibitem{Carena:1994bv} 
  M.~S.~Carena, M.~Olechowski, S.~Pokorski and C.~E.~M.~Wagner,
  Nucl.\ Phys.\ B {\bf 426}, 269 (1994)
  [hep-ph/9402253].

\bibitem{Carena:1998gk} 
  M.~S.~Carena, S.~Mrenna and C.~E.~M.~Wagner,
  Phys.\ Rev.\ D {\bf 60}, 075010 (1999)
  [hep-ph/9808312].

\bibitem{Eberl:1999he} 
  H.~Eberl, K.~Hidaka, S.~Kraml, W.~Majerotto and Y.~Yamada,
  Phys.\ Rev.\ D {\bf 62}, 055006 (2000)
  [hep-ph/9912463].

\bibitem{Carena:1999py} 
  M.~S.~Carena, D.~Garcia, U.~Nierste and C.~E.~M.~Wagner,
  Nucl.\ Phys.\ B {\bf 577}, 88 (2000)
  [hep-ph/9912516].

\bibitem{Hofer:2009xb} 
  L.~Hofer, U.~Nierste and D.~Scherer,
  JHEP {\bf 0910}, 081 (2009)
  [arXiv:0907.5408 [hep-ph]].

\bibitem{Noth:2008tw} 
  D.~Noth and M.~Spira,
  Phys.\ Rev.\ Lett.\  {\bf 101}, 181801 (2008)
  [arXiv:0808.0087 [hep-ph]].

\bibitem{Carena:2001bg}
  M.~S.~Carena, H.~E.~Haber, H.~E.~Logan and S.~Mrenna,
  Phys.\ Rev.\ D {\bf 65} (2002) 055005
   [Erratum-ibid.\ D {\bf 65} (2002) 099902]
  [hep-ph/0106116].

\bibitem{Arvanitaki:2011ck} 
  A.~Arvanitaki and G.~Villadoro,
  JHEP {\bf 1202}, 144 (2012)
  [arXiv:1112.4835 [hep-ph]].

\bibitem{Reece:2012gi} 
  M.~Reece,
  New J.\ Phys.\  {\bf 15}, 043003 (2013)
  [arXiv:1208.1765 [hep-ph]].

\bibitem{Bagnaschi:2014zla} 
  E.~Bagnaschi, R.~V.~Harlander, S.~Liebler, H.~Mantler, P.~Slavich and A.~Vicini,
  JHEP {\bf 1406}, 167 (2014)
  [arXiv:1404.0327 [hep-ph]].

\bibitem{Heinemeyer:2000fa}
  S.~Heinemeyer, W.~Hollik and G.~Weiglein,
  Eur.\ Phys.\ J.\ C {\bf 16} (2000) 139
  [hep-ph/0003022].


\bibitem{Cahill-Rowley:2014wba}
  M.~Cahill-Rowley, J.~Hewett, A.~Ismail and T.~Rizzo,
  Phys.\ Rev.\ D {\bf 90} (2014) 9,  095017
  [arXiv:1407.7021 [hep-ph]].

\bibitem{Babu:1999hn} 
  K.~S.~Babu and C.~F.~Kolda,
  Phys.\ Rev.\ Lett.\  {\bf 84}, 228 (2000)
  [hep-ph/9909476].

\bibitem{Choudhury:1998ze} 
  S.~R.~Choudhury and N.~Gaur,
  Phys.\ Lett.\ B {\bf 451}, 86 (1999)
  [hep-ph/9810307].

\bibitem{Buchalla:1993bv} 
  G.~Buchalla and A.~J.~Buras,
  Nucl.\ Phys.\ B {\bf 400}, 225 (1993).

\bibitem{Misiak:1999yg} 
  M.~Misiak and J.~Urban,
  Phys.\ Lett.\ B {\bf 451}, 161 (1999)
  [hep-ph/9901278].

\bibitem{Bobeth:2001sq} 
  C.~Bobeth, T.~Ewerth, F.~Kruger and J.~Urban,
  Phys.\ Rev.\ D {\bf 64}, 074014 (2001)
  [hep-ph/0104284].

\bibitem{Bobeth:2001jm} 
  C.~Bobeth, A.~J.~Buras, F.~Kruger and J.~Urban,
  Nucl.\ Phys.\ B {\bf 630}, 87 (2002)
  [hep-ph/0112305].
  
\bibitem{Bobeth:2013uxa}
  C.~Bobeth, M.~Gorbahn, T.~Hermann, M.~Misiak, E.~Stamou and M.~Steinhauser,
  Phys.\ Rev.\ Lett.\  {\bf 112} (2014) 101801
  [arXiv:1311.0903 [hep-ph]].

\bibitem{CMS:2014xfa}
  V.~Khachatryan {\it et al.}  [CMS and LHCb Collaborations],
  arXiv:1411.4413 [hep-ex].

\bibitem{Amhis:2014hma} 
  Y.~Amhis {\it et al.}  [Heavy Flavor Averaging Group (HFAG) Collaboration],
  arXiv:1412.7515 [hep-ex].

\bibitem{Misiak:2006zs} 
  M.~Misiak, H.~M.~Asatrian, K.~Bieri, M.~Czakon, A.~Czarnecki, T.~Ewerth, A.~Ferroglia and P.~Gambino {\it et al.},
  Phys.\ Rev.\ Lett.\  {\bf 98}, 022002 (2007)
  [hep-ph/0609232].

\bibitem{Mahmoudi:2008tp}
  F.~Mahmoudi,
  Comput.\ Phys.\ Commun.\  {\bf 180} (2009) 1579
  [arXiv:0808.3144 [hep-ph]].

\bibitem{Altmannshofer:2012ks}
  W.~Altmannshofer, M.~Carena, N.~R.~Shah and F.~Yu,
  JHEP {\bf 1301} (2013) 160
  [arXiv:1211.1976 [hep-ph]].

\bibitem{Frere:1983ag} 
  J.~M.~Frere, D.~R.~T.~Jones and S.~Raby,
  Nucl.\ Phys.\ B {\bf 222}, 11 (1983).

\bibitem{Gunion:1987qv} 
  J.~F.~Gunion, H.~E.~Haber and M.~Sher,
  Nucl.\ Phys.\ B {\bf 306}, 1 (1988).

\bibitem{Casas:1995pd} 
  J.~A.~Casas, A.~Lleyda and C.~Munoz,
  Nucl.\ Phys.\ B {\bf 471}, 3 (1996)
  [hep-ph/9507294].

\bibitem{Kusenko:1996jn} 
  A.~Kusenko, P.~Langacker and G.~Segre,
  Phys.\ Rev.\ D {\bf 54}, 5824 (1996)
  [hep-ph/9602414].

\bibitem{Camargo-Molina:2013sta} 
  J.~E.~Camargo-Molina, B.~O'Leary, W.~Porod and F.~Staub,
  JHEP {\bf 1312}, 103 (2013)
  [arXiv:1309.7212 [hep-ph]].

\bibitem{Chowdhury:2013dka} 
  D.~Chowdhury, R.~M.~Godbole, K.~A.~Mohan and S.~K.~Vempati,
  JHEP {\bf 1402}, 110 (2014)
  [arXiv:1310.1932 [hep-ph]].

\bibitem{Blinov:2013fta} 
  N.~Blinov and D.~E.~Morrissey,
  JHEP {\bf 1403}, 106 (2014)
  [arXiv:1310.4174 [hep-ph]].

\bibitem{Camargo-Molina:2014pwa} 
  J.~E.~Camargo-Molina, B.~Garbrecht, B.~O'Leary, W.~Porod and F.~Staub,
  Phys.\ Lett.\ B {\bf 737}, 156 (2014)
  [arXiv:1405.7376 [hep-ph]].

\bibitem{Coleman:1977py}
  S.~R.~Coleman,
  Phys.\ Rev.\ D {\bf 15} (1977) 2929
   [Erratum-ibid.\ D {\bf 16} (1977) 1248].

\bibitem{Callan:1977pt}
  C.~G.~Callan, Jr. and S.~R.~Coleman,
  Phys.\ Rev.\ D {\bf 16} (1977) 1762.

\bibitem{Wainwright:2011kj}
  C.~L.~Wainwright,
  Comput.\ Phys.\ Commun.\  {\bf 183} (2012) 2006
  [arXiv:1109.4189 [hep-ph]].

\bibitem{Williams:2011bu} 
  K.~E.~Williams, H.~Rzehak and G.~Weiglein,
  Eur.\ Phys.\ J.\ C {\bf 71}, 1669 (2011)
  [arXiv:1103.1335 [hep-ph]].

\bibitem{Peskin:2013xra}
  M.~E.~Peskin,
  arXiv:1312.4974 [hep-ph].

\end{thebibliography}
\end{document}